*Article*

# Optimizing a multi-state cold-standby system with multiple vacations in the repair and loss of units

**Juan Eloy Ruiz-Castro [1,*]**

[1] Department of Statistics and O.R. and Math Institute (IMAG), University of Granada, 18071 Granada, Spain
* Correspondence: jeloy@ugr.es

**Abstract:** A complex multi-state redundant system with preventive maintenance subject to multiple events is considered. The online unit can undergo several types of failures: internal and those provoked by external shocks. Multiple degradation levels are assumed so as internal and external. Degradation levels are observed by random inspections and if they are major, the unit goes to repair facility where preventive maintenance is carried out. This repair facility is composed of a single repairperson governed by a multiple vacation policy. This policy is set up according to the operational number of units. Two types of task can be performed by the repairperson, corrective repair and preventive maintenance. The times embedded in the system are phase type distributed and the model is built by using Markovian Arrival Processes with marked arrivals. Multiple performance measures besides of the transient and stationary distribution are worked out through matrix-analytic methods. This methodology enables us to express the main results and the global development in a matrix-algorithmic form. To optimize the model costs and rewards are included. A numerical example shows the versatility of the model.

## 1. Introduction

Redundant systems and preventive maintenance are of fundamental importance in ensuring reliability, preventing system failures and reducing costs. These questions, therefore, are of considerable research interest.

The occurrence of total, unexpected system failure can provoke severe damage and major financial loss. To avoid such an outcome, various reliability-enhancing methods can be applied, chief among which are redundancy and preventive maintenance. In this respect, cold, hot and warm redundant standby and *k*-out-of-*n* systems have been proposed. Among researchers who have addressed these questions, [1] considered an optimal standby element sequencing problem (SESP) for 1-out-of-N: G heterogeneous warm-standby systems, while [2] constructed a multi-value decision diagram with which to analyse a demand-based warm standby system. In related papers, [3] considered preventive maintenance for items operating in a random environment subjected to a shock Poisson process, [4] evaluated the probability of mission success given an arbitrary redundancy level, and [5] analysed the behaviour of a two-unit standby redundant system.

Preventive maintenance enhances system reliability and performance, reduces costs, for both repairable and non-repairable systems, and decreases the probability of sudden equipment failure. Various maintenance systems were studied by [6], and [7] developed a new model for the hybrid preventive maintenance of systems with partially observable degradation. [8] modelled the (time-consuming) procedure of task transfer, in an event transition-based reliability analysis of standby systems in which preventive replacements are performed according to a predetermined schedule. The aim of this approach is to optimise preventive replacement scheduling and hence to maximise reliability. In another approach to this situation, [9] discussed a preventive maintenance policy for a single-unit system subject to failure by internal deterioration and/or sudden shock, according to a non-homogeneous Poisson process whereby the process of internal failure is partitioned into two stages.

Complex systems that have a finite number of performance levels and various failure modes, each producing different effects on system performance, are termed multi-state systems (MSS). This concept was first discussed by [10] and has since been developed extensively. For example, [11] described various MSS measures and considered problems of MSS optimisation, and [12] conducted a comprehensive analysis of the question.

One of the main problems encountered with multi-state complex models is the existence of intractable expressions for their modelling and/or of difficulties in their interpretation. In this respect, matrix-analytic methods are a valuable means of analysing complex systems, preserving the Markovian structure and obtaining manageable results. This approach is usually based on two elements – phase-type distributions (PHD) and Markovian arrival processes (MAP) – which enable the results to be expressed and complex systems modelled in an algorithmic, computational form. PHD were first introduced and detailed by [13]. MAP is a counting process in which PH distributions play an important role. This method was described by [14] and comprehensively reviewed by [15] and [16]. A special case is that of the MAP with marked arrivals (MMAP), which enables us to count different types of arrivals. Moreover, the arrival probabilities of events, for the discrete case, can be customised for different situations. MAP and MMAP theory was further developed by [16].

Many multi-state reliability systems, over time, are subject to events such as repairable or non-repairable failure, inspections or external shocks. These systems can be modelled using appropriate Markov processes, i.e. PHD and MAP ([17], (18]). In parallel, unitary complex systems subject to multiple events have been discussed by [19] and (20]. Matrix algorithmic methods have also been used to model multi-state complex redundant systems. [21] developed a k-out-of-n: G system, in which the units are subject to repairable and/or non-repairable failure and receive random inspections. In this system, the potential loss of units is included; thus, when a non-repairable failure occurs, the unit is removed and the system continues to be operational. In the context of complex models, a repair facility with a single repairperson is usually assumed. Thus, [22] and [23] analyse redundant complex systems with a general number of repairpersons and the potential loss of units, determining the optimum number of repairpersons in each case.

In brief, redundancy and preventive maintenance are incorporated into complex systems in order to enhance their reliability, and must also be included in the modelling of such systems. In theory, a unit is repaired either immediately after failure if the system is unitary or when the element in next in line in the repair facility queue. However, this might not be the case in a real scenario. For example, a failed unit might not be repaired immediately in a small or medium-sized firm that cannot afford to employ a full-time repairperson. Furthermore, when there is no failed unit to be attended in the repair facility, what should a repairperson do? Instead of remaining idle during this period, the repairperson may take a 'vacation' and/or use the time to do other work, thus optimising resources and reducing costs. A repairperson is on vacation when absent from the repair facility, whether or not it is empty. The economic implications of this situation should be considered, taking into account that the vacation policy applied might impact both on performance and also on economic rewards/costs. In studies of this question, two time points are of particular importance: the start and end times of the vacation. Moreover, the services provided may be exhaustive or non-exhaustive. In the first case, the repairperson cannot be on vacation when the repair facility is not empty, but in the second, even if an item has been sent to the repair facility, the repairperson may be on vacation. Another possibility that must be considered is that of interruption, i.e. the repairperson may take a vacation while a unit is being repaired. The vacation end time determines when the repairperson resumes work. Finally, depending on the maintenance system adopted, the vacation may occupy a single period of time or multiple periods.

Vacation policies have been considered in queuing theory and in reliability analysis, among other areas. Thus, [24] provided a wide-ranging analysis of vacation system models and [25] examined the application of two vacation policies (one single and the other multiple) in a repairable system. [26] developed a reliability system with multiple, but finite, vacation periods and [27] analysed the reliability of a two-unit cold standby system with a single repairperson, entitled to take a vacation.

Vacation periods have also been considered for systems governed by a Markov model. In this respect, [28] presented the case of an exhaustive vacation policy, whereby the repairperson could only take a vacation when the repair facility was empty. Under the Markovian modelling described by [29], the repairperson could take a vacation if there were no failed units in need of repair, but had to return as soon as any unit failed. In another approach, [30] modelled a k-out-of-n system with a single repairperson, assuming a phase-type distribution for the vacation time and an exponential distribution for the lifetime of the units. In this system, the repairperson could take a vacation whenever there was no failed component in the system. On return, the repairperson might or might not encounter failed components waiting for repair. In the second case, the repairperson would remain within the repair facility, idle, until a failed component arrived. Finally, [31] modelled a multi-state complex system subject to multiple events and where preventive maintenance was applied. In this case, the repairperson had various duties and, moreover, was entitled to take a vacation.

In the present study, we model a cold standby system with the potential loss of units. The system evolves in discrete time; the online unit is multi-state and subject to internal failure, repairable or otherwise, to external shocks with diverse consequences, and to random inspection. When a non-repairable failure occurs, the faulty unit is removed and the system continues working with one unit less. An external shock may provoke any of the following consequences: degraded system performance, a repairable failure of the online unit or its total (non-repairable) failure. Damage to the internal performance of the online unit may be minor or major. During system inspection, the internal status of the online unit is observed. If major damage is present, the faulty unit is sent to the repair facility for preventive maintenance. According to the case presented, the repairperson may perform corrective repair or preventive maintenance. The complexity of the system is determined as follows. In modelling the system, the vacation policy employed in the repair facility is determined by the number of operational units included. A general number $R$ of operational units is considered. If the repairperson returns from a vacation period and there are fewer than $R$ operational units, the repairperson must then remain in the facility. Otherwise, a new vacation period begins. As the system is subject to a loss of units, when there are fewer than $R$ units in the system, the repairperson must remain in the facility while this situation persists. The times embedded are PH distributed and a MMAP is constructed to model the system. In modelling this system, the following measures are calculated: availability, reliability and expected times (in both transient and stationary regimes). Rewards and costs are incorporated, and a numerical optimisation is performed to determine the optimum threshold $R$ and to decide whether preventive maintenance is profitable or not.

The rest of this paper is organised as follows. In Section 2, we describe the system to be modelled, after which we present the corresponding MMAP in Section 3. In Section 4, we detail the measures applied to the transient and stationary regimes, and calculate the transient and stationary distributions. The latter is obtained both algorithmically and computationally. The system costs, rewards and associated measures are then derived in Section 5. Taking advantage of the favourable properties of PHD and MMAP, the study findings are obtained in a matrix algorithmic form. Section 6 presents a numerical example, including an optimisation exercise. Finally, the main conclusions drawn are summarised in Section 7.

## 2. Assumptions of the system: the state space

A cold standby system composed of $n$ units initially is assumed. One unit is online and the others are waiting on standby without degrading. The online unit is multi-state where the internal performance is partitioned into major and minor states. It is subject to multiple events. This can suffer internal failures, repairable or not, and external shocks. Each external shock can provoke three different consequences: total failure (non-repairable), modification in the internal behaviour or even an internal repairable or non-repairable failure. When a repairable failure occurs, the unit goes to the repair facility for corrective repair~~.~~. The corrective repair time distribution is PH. The repair facility is composed of one repairperson who can take vacations. As it has been mentioned above, the internal performance of the online unit is partitioned into major and minor states. A major state is a state from where the online unit has a greater risk of suffering a failure. To avoid serious damage and major financial losses ~~Randomly,~~ random inspections are carried out. The inspector observes the online unit and if this one is operational in major damage, the unit goes to the repair facility for preventive maintenance. Preventive maintenance time is also PH distributed. When the online unit undergoes a failure, one cold standby occupies the online place, if any. The new online unit will start executing from the initial distribution of the internal performance because after repairing or preventive maintenance the unit is as good as new. Also, ~~The~~ the system is subject to loss of units. After a non-repairable failure the unit is removed and the system continues working until there are no units in the system. If only one unit is in the system and a non-repairable failure occurs, the system is restarted.

One repairperson can be in the repair facility who can develop two different tasks: corrective repair and preventive maintenance. To optimise the system, the repairperson is allowed to take vacations, for a random duration, according to certain criteria.

Initially, all units are operational and the repairperson is on vacation. After returning, a new vacation begins if there are $R$ or more operational units in the system. Equivalently, if there are $k–R+1=N$ or more failed units needing to be repaired, where $k$ is the number of units in the system, $k = 1,..., n$, the repairperson must remain in the repair facility.

After finishing a repair, the repairperson begins a new period of vacation if $R$ units are then operational. As the system can lose units, the repairperson must always remain in the facility (or interrupt the vacation to return) when fewer than $R$ units are in the system.

The following Section 2.1 specifies the assumptions of the system.

### *2.1. The assumptions*

The system follows the next assumptions.

Assumption 1. The internal performance time is PH distributed with representation $(\boldsymbol{\alpha}, \mathbf{T})$, with order *m* (number of internal stages). The internal failure probability depends on the states. The column vectors $\mathbf{T}_r^0$ and $\mathbf{T}_{nr}^0$ contains the probabilities of repairable and non-repairable failures respectively.

Assumption 2. The internal performance of the online unit is multi-state where the $n_1$ first units are minor and the rest major according to damage.

Assumption 3. The external events occur according to a PH-renewal process where the time between two consecutive shocks is a PH distribution with representation $(\boldsymbol{\gamma}, \mathbf{L})$, with order *t*.

Assumption 4. An external shock can provoke a total non-repairable failure of the online unit with a probability equal to $\omega^0$.

Assumption 5. After an external shock the internal performance state can undergo a modification. This modification between any two internal states occurs according to the transition probability matrix **W**. The column vectors $\mathbf{W}_r^0$ and $\mathbf{W}_{nr}^0$ contains the probabilities of repairable and non-repairable failures respectively provoked by an external shock.

Assumption 6. The time between two consecutive random inspections is PH distributed with representation $(\boldsymbol{\eta}, \mathbf{M})$, with order $\varepsilon$.

Assumption 7. The vacation time is distributed following a PH distribution with representation $(\mathbf{v}, \mathbf{V})$, with order $\upsilon$.

Assumption 8. The corrective repair time is PH distributed with representation $(\boldsymbol{\beta}_1, \mathbf{S}_1)$, with order $z_1$.

Assumption 9. The preventive maintenance time is PH distributed with representation $(\boldsymbol{\beta}_2, \mathbf{S}_2)$, with order $z_2$.

The behaviour of the system is shown in Figure 1, for inspection and repairable failure, Figure 2 for non-repairable failure, and Figure 3 for the vacation policy.

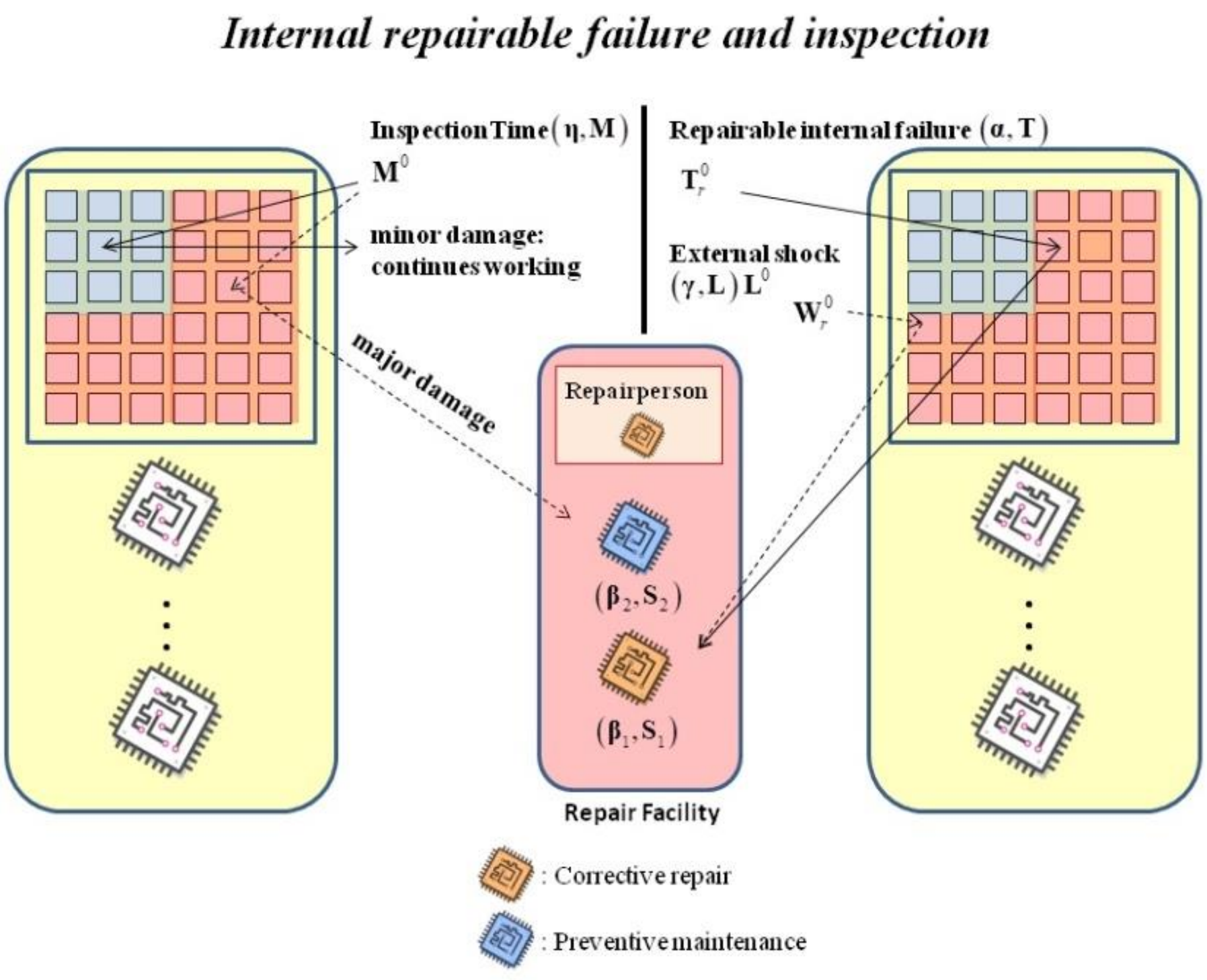

**Figure 1.** Internal repairable failure and inspection in the system

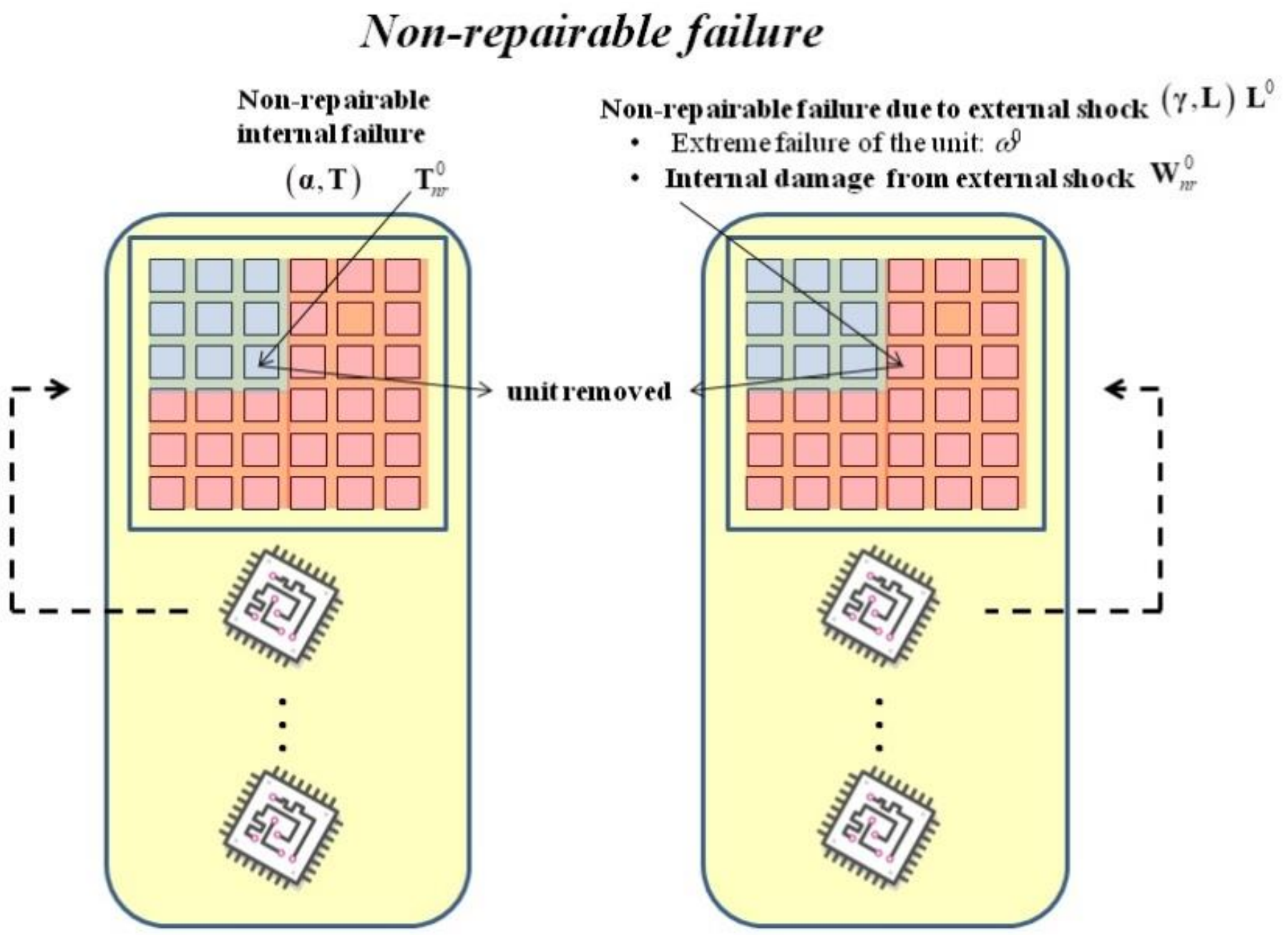

**Figure 2.** Non-repairable failure in the system

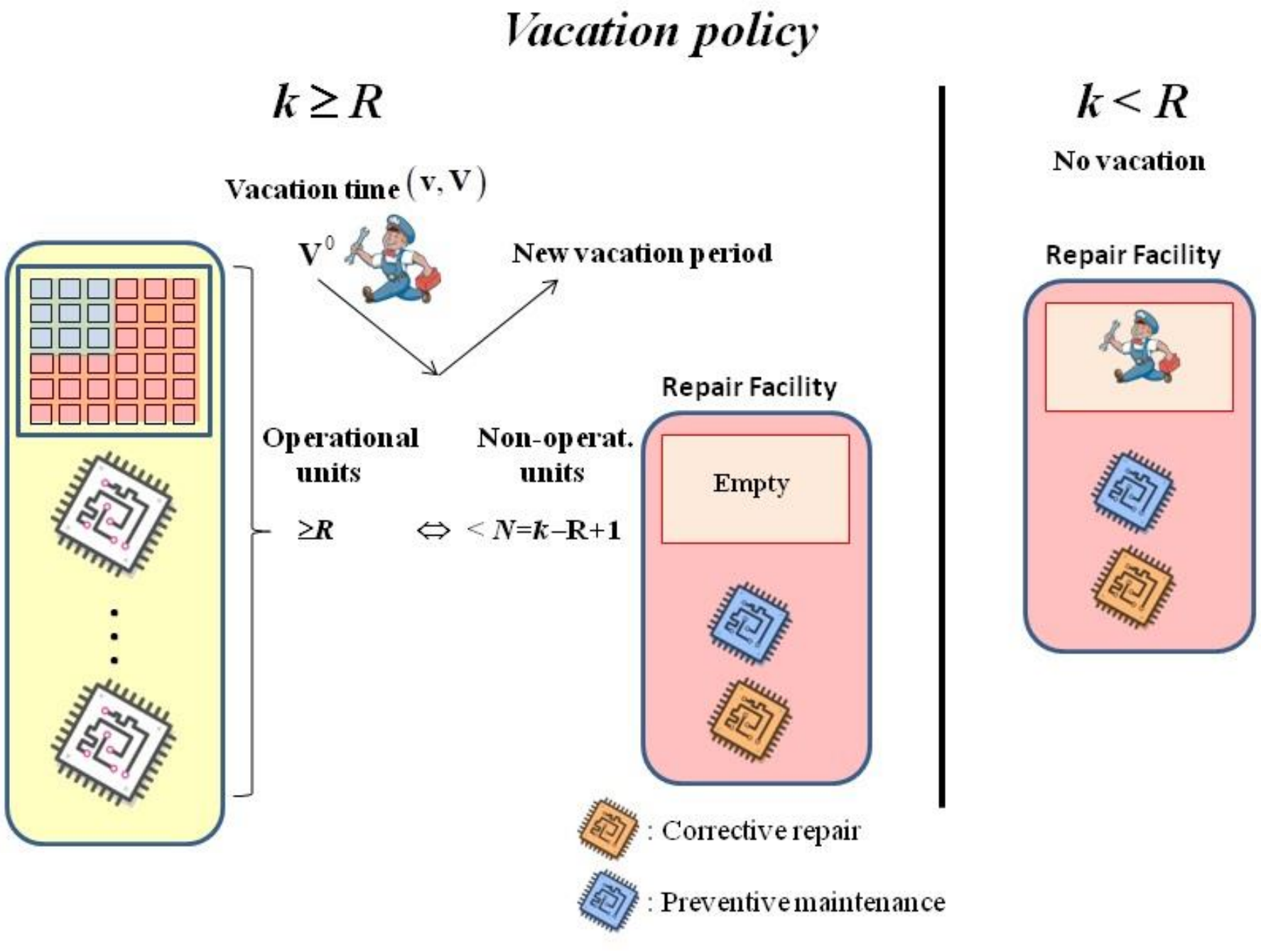

**Figure 3.** Vacation policy in the system

## 3. Modelling the system. The Markovian Arrival Process with marked arrivals

The system is governed by a Markov process vector in discrete time. In this section the state space is described and to model the proposed complex system, the behaviour of the online unit and of the repair facility is developed separately.

### 3.1. The state-space

The state-space is composed of macro-states and it is denoted by $S=\left\{\mathbf{U}^{n},\mathbf{U}^{n-1},\ldots,\mathbf{U}^{1}\right\}$, where $\mathbf{U}^{k}$ containing the phases when there are $k$ units in the system. In turn, these macro-states are partitioned as follows

$$\mathbf{U}^{k}=\left\{\mathbf{E}_{0}^{k,v},\mathbf{E}_{1}^{k,v},...,\mathbf{E}_{N-1}^{k,v},\mathbf{E}_{N}^{k,v},\mathbf{E}_{N+1}^{k,v},\ldots,\mathbf{E}_{k}^{k,v},\mathbf{E}_{N}^{k,nv},\mathbf{E}_{N+1}^{k,nv},\ldots,,\mathbf{E}_{k}^{k,nv}\right\};\ k\geq R$$

$$\mathbf{U}^k = \left\{\mathbf{E}_0^{k,nv}, \mathbf{E}_1^{k,nv}, \ldots, \mathbf{E}_k^{k,nv}\right\};\ k < R$$

where $\mathbf{E}_s^{k,x}$ contains the phases when there are $k$ units in the system and $s$ of them are in the repair facility and the superscript $x$ indicates if the repairperson in on vacation ($v$) or not ($nv$). Initially the repairperson begins to operate the first time that he comes back from holidays and the system has at least $N = k–R+1$ units in the repair facility. He remains working until $N–1$ units are in the repair facility. At this moment the repairperson goes on vacation. In any case, the order of the units in the repair facility has to be saved in memory, and there are two types of repair, corrective and preventive maintenance. For this reason, the macro-state $\mathbf{E}_s^{k,x}$ is composed of the first level of macro-states $\mathbf{E}_{i_1,\ldots,i_s}^{k,x}$.

These macro-states contain the phases when there are k units in the system, with $s$ of them in the repair facility, and the type of repair is given by the ordered sequence $i_1, \ldots, i_s$. The values of $i_l$ are equal to 1 or 2 if the unit is in corrective repair or preventive maintenance, respectively.

When the number of units in the system is $R–1$ units, then the repairperson occupies his place work immediately. The inspection time is restarted each time that one unit occupies the online place.

- For $k = 1, \ldots, R–1$

$$\mathbf{E}_0^{k,nv} = \left\{(k,0;i,j,u);i=1,\ldots,m,\ j=1,\ldots,t,\ u=1,\ldots,\varepsilon\right\}$$

$$\mathbf{E}_s^{k,nv} = \left\{\mathbf{E}_{i_1,\ldots,i_s}^{k,nv};i_l=1,2;l=1,\ldots,s\right\} \text{ for } s = 1,\ldots,k \text{ where}$$

$$\mathbf{E}_{i_1,\ldots,i_s}^{k,nv} = \left\{(k,s;i,j,u,r);i=1,\ldots,m,\ j=1,\ldots,t,u=1,\ldots,\varepsilon,r=1,\ldots,z_{i_1}\right\} \text{ for } s < k$$

$$\mathbf{E}_{i_1,\ldots,i_k}^{k,nv} = \left\{(k,k;j,r);\ j=1,\ldots,t,u=1,\ldots,\varepsilon,r=1,\ldots,z_{i_1}\right\}$$

- For $k = N, \ldots, n$

$$\mathbf{E}_0^{k,v} = \left\{(k,0;i,j,u,v);i=1,\ldots,m,\ j=1,\ldots,t,\ u=1,\ldots,\varepsilon,v=1,\ldots,\upsilon\right\}$$

$$\mathbf{E}_s^{k,v} = \left\{\mathbf{E}_{i_1,\ldots,i_s}^{k,v};i_l=1,2;l=1,\ldots,s\right\} \text{ for } s = 1,\ldots,k \text{ where}$$

$$\mathbf{E}_{i_1,\ldots,i_s}^{k,v} = \left\{(k,s;i,j,u,v);i=1,\ldots,m,\ j=1,\ldots,t,\ u=1,\ldots,\varepsilon,v=1,\ldots,\upsilon\right\} \text{ for } s < k$$

$$\mathbf{E}_{i_1,\ldots,i_k}^{k,v} = \left\{(k,k;j,u,v);\ j=1,\ldots,t,\ u=1,\ldots,\varepsilon,v=1,\ldots,\upsilon\right\}$$

$$\mathbf{E}_s^{k,nv} = \left\{\mathbf{E}_{i_1,\ldots,i_s}^{k,nv};i_l=1,2;l=1,\ldots,s\right\} \text{ for } s = N,\ldots,k \text{ where}$$

$$\mathbf{E}_{i_1,\ldots,i_s}^{k,nv} = \left\{(k,s;i,j,u,r);i=1,\ldots,m,\ j=1,\ldots,t,u=1,\ldots,\varepsilon,r=1,\ldots,z_{i_1}\right\} \text{ for } s < k$$

$$\mathbf{E}_{i_1,\ldots,i_k}^{k,nv} = \left\{(k,k;j,u,r);\ j=1,\ldots,t,u=1,\ldots,\varepsilon,r=1,\ldots,z_{i_1}\right\}$$

The phase $(k,s;i,j,u,m,r)$ indicates that there are $k$ units in the system, with $s$ in the repair facility; the internal performance of the online unit is in state $i$, the external shock time is in state $j$, the cumulative damage caused by external shocks is given by $u$, $m$ is the current phase of the inspection time and $r$ is the corrective repair/preventive maintenance phase for the unit currently being attended in the repair facility. If the repairperson is taking a vacation, the phase is indicated by $v$.

The order of these macro-states is the following:

$$o_{\mathbf{E}_0^{k,nv}} = m\cdot t\cdot\varepsilon \quad ; \quad s<k,\ o_{\mathbf{E}_s^{k,nv}} = m\cdot t\cdot\varepsilon\cdot(z_1+z_2)2^{s-1}\ ;\ s=k,\ o_{\mathbf{E}_k^{k,nv}} = t\cdot(z_1+z_2)2^{k-1}$$

$$o_{\mathbf{E}_0^{k,v}} = m\cdot t\cdot\varepsilon \quad ; \quad s<k,\ o_{\mathbf{E}_s^{k,v}} = m\cdot t\cdot\varepsilon\cdot 2^{s}\ ;\ s=k,\ o_{\mathbf{E}_k^{k,v}} = t\cdot 2^{k-1}$$

### *3.2. Modelling the online unit*

The online unit can undergo different types of events at any time. These ones are noted and defined as:

*A*: Internal repairable failure

*B*: Major revision

*C*: Non-repairable failure

*O*: No events

Two of them are described below, the rest are given in Appendix A.

The elements of auxiliary matrices $\mathbf{U}_1$ and $\mathbf{U}_2$ are defined as

$$\mathbf{U}_1(i,j) = \begin{cases}1 & ;\ i=j;i=1,\ldots,n_1 \\ 0 & ;\ \text{otherwise}\end{cases};\ \mathbf{U}_2(i,j) = \begin{cases}1 & ;\ i=j;i=n_1+1,\ldots,n \\ 0 & ;\ \text{otherwise}\end{cases}$$

Throughout this work the symbol ⊗ denotes the Kronecker product and given a matrix **A**, we denote as $\mathbf{A}^0$ to the column vector $\mathbf{A}^0 = \mathbf{e} - \mathbf{Ae}$, being **e** a column vector of ones with appropriate order.

*No events at a certain time (O)*

We assume that the online unit is operational and at this time it continues working. It occurs because of different situations:

- The internal performance continues in the same phase or changes to another, equally operational state. There is no external shock ($\mathbf{T}\otimes\mathbf{L}$), and no inspection takes place (**M**). The matrix that governs this transition for the online unit is given by $\mathbf{T}\otimes\mathbf{L}\otimes\mathbf{M}$.
- The online undergoes an external shock but total failure does not occur($\mathbf{L}^0\boldsymbol{\gamma}\left(1-\omega^0\right)$). This external shock might modify the internal performance but does not produce internal failure ($\mathbf{TW}$). No inspection takes place (**M**). The matrix is $\left(\mathbf{TW}\otimes\mathbf{L}^0\boldsymbol{\gamma}\left(1-\omega^0\right)\right)\otimes\mathbf{M}$.
- An inspection takes place and the time preceding the next one begins ($\mathbf{M}^0\boldsymbol{\eta}$). The inspector observes that the online unit does not need preventive maintenance and no external shock occurs ($\mathbf{U}_1\mathbf{T}\otimes\mathbf{L}$). The matrix is $\mathbf{U}_1\mathbf{T}\otimes\mathbf{L}\otimes\mathbf{M}^0\boldsymbol{\eta}$.
- An inspection takes place and the time preceding the next one begins ($\mathbf{M}^0\boldsymbol{\eta}$). Also, one external shock takes place without total failure ($\mathbf{L}^0\boldsymbol{\gamma}\left(1-\omega^0\right)$). This shock provokes a change in the internal performance without failure and the inspection observes minor damage ($\mathbf{U}_1\mathbf{TW}$). This matrix is $\left(\mathbf{U}_1\mathbf{TW}\otimes\mathbf{L}^0\boldsymbol{\gamma}\left(1-\omega^0\right)\right)\otimes\mathbf{M}^0\boldsymbol{\eta}$.

Therefore, the matrix that governs this transition for the online unit is given by

$$\mathbf{H}_O = \left(\mathbf{T}\otimes\mathbf{L}+\mathbf{TW}\otimes\mathbf{L}^0\boldsymbol{\gamma}\left(1-\omega^0\right)\right)\otimes\mathbf{M}+\left(\mathbf{U}_1\mathbf{T}\otimes\mathbf{L}+\mathbf{U}_1\mathbf{TW}\otimes\mathbf{L}^0\boldsymbol{\gamma}\left(1-\omega^0\right)\right)\otimes\mathbf{M}^0\boldsymbol{\eta}.$$

*Non-repairable failure (C)*

The online unit is assumed to be operational and at the next time point a non-repairable failure occurs, because:

- An internal non-repairable failure occurs with no external shock, $\mathbf{T}_{nr}^0\boldsymbol{\alpha}\otimes\mathbf{L}$.
- An external shock occurs, but does not provoke total failure. This shock provokes a non-repairable internal failure or, irrespective of the shock, the online unit may experience a non-repairable internal failure. The matrix is $\left(\mathbf{T}_{nr}^0+\mathbf{TW}_{nr}^0\right)\boldsymbol{\alpha}\otimes\mathbf{L}^0\boldsymbol{\gamma}\left(1-\omega^0\right)$.
- An external shock provokes total failure. In this case the internal behaviour is irrelevant. The matrix is $\mathbf{e}\boldsymbol{\alpha}\otimes\mathbf{L}^0\boldsymbol{\gamma}\omega^0$.

This transition is independent of the inspection time. After the online unit experiences a non-repairable failure, the online place is occupied by a substitute, identical unit. Then, the matrix is given by

$$\mathbf{H}_C = \left[\mathbf{T}_{nr}^0\boldsymbol{\alpha}\otimes\mathbf{L}+\left(\mathbf{T}_{nr}^0+\mathbf{TW}_{nr}^0\right)\boldsymbol{\alpha}\otimes\mathbf{L}^0\boldsymbol{\gamma}\left(1-\omega^0\right)+\mathbf{e}\boldsymbol{\alpha}\otimes\mathbf{L}^0\boldsymbol{\gamma}\omega^0\right]\otimes\mathbf{e}\boldsymbol{\eta}.$$

If only one unit is operational and online (i.e. all others are under repair), this unit experiences a non-repairable failure and no repair occurs, no immediate substitution can be made and therefore the system does not restart. The matrix is given by

$$\mathbf{H'}_C = \left[\mathbf{T}_{nr}^0\otimes\mathbf{L}+\left(\mathbf{T}_{nr}^0+\mathbf{TW}_{nr}^0\right)\otimes\mathbf{L}^0\boldsymbol{\gamma}\left(1-\omega^0\right)+\mathbf{e}\otimes\mathbf{L}^0\boldsymbol{\gamma}\omega^0\right]\otimes\mathbf{e}.$$

### *3.3. The Markovian Arrival Process with marked arrivals (MMAP)*

The behaviour of the system is governed by a MMAP. The representation of this MMAP is given from the types of events shown below:

*A*: Internal repairable failure (default without D)
*B*: Major revision (default without D)
*C*: Non-repairable failure (default without D)
*D*: The repairperson resumes to work (default without A, B, C)

*AD*: Internal repairable failure and the repairperson resumes to work
*BD*: Major revision and the repairperson resumes to work
*CD*: Non-repairable failure and the repairperson resumes to work
*NS*: New system
*O*: No events
The representation of the MMAP is $\left(\mathbf{D}^{O},\mathbf{D}^{A},\mathbf{D}^{B},\mathbf{D}^{C},\mathbf{D}^{D},\mathbf{D}^{AD},\mathbf{D}^{BD},\mathbf{D}^{CD},\mathbf{D}^{NS}\right)$ .
The transition probability matrix associated to the embedded Markov chain from the MMAP is given by $\mathbf{D}=\sum_{Y}\mathbf{D}^{Y}$
.

Two matrices $\mathbf{D}^{Y}$ are described in the next section. The rest are given in Appendix B.

### *3.3.1. The matrices* $\mathbf{D}^{A}$ *and* $\mathbf{D}^{B}$

The matrices $\mathbf{D}^{A}$ and $\mathbf{D}^{B}$ govern the transition when a repairable failure or a major inspection takes place, respectively. These matrices are composed of matrix blocks that contain the transitions between macro-states $\mathbf{U}^{k}$. This is a diagonal matrix block given that the number of units in the system does not change in this transition. The matrix $\mathbf{D}_{k}^{Y}$ contains the transition probabilities when there are *k* units in the system and the event *Y* occurs for *Y* = *A* or *B* and *k*=1,…,*n*. Then,

$$\mathbf{D}^{Y}=\begin{pmatrix}\mathbf{D}_{n}^{Y} & & & & \\ & \mathbf{D}_{n-1}^{Y} & & & \\ & & \mathbf{D}_{n-2}^{Y} & & \\ & & & \ddots & \\ & & & & \mathbf{D}_{1}^{Y}\end{pmatrix}\quad \text{for } Y=A,\ B.$$

These blocks are composed of blocks again.

- If the number of units is less than *R*–1, the repairperson is always in his workplace. Then, for *k* =1,…, *R*–1

$$\mathbf{D}_{k}^{Y}=\begin{array}{c} E_{0}^{k,nv} \\ E_{1}^{k,nv} \\ \vdots \\ E_{k-1}^{k,vn} \\ E_{k}^{k,nv}\end{array}\overset{\begin{array}{ccccc} E_{0}^{k,nv} & E_{1}^{k,nv} & E_{2}^{k,nv} & E_{k-1}^{k,nv} & E_{k}^{k,nv}\end{array}}{\begin{pmatrix}\mathbf{0} & \mathbf{D}_{01}^{Y,k,nv} & & & \\ & \mathbf{D}_{11}^{Y,k,nv} & \mathbf{D}_{12}^{Y,k,nv} & & \\ & & \ddots & \ddots & \\ & & & \mathbf{D}_{k-1,k-1}^{Y,k,nv} & \mathbf{D}_{k-1,k}^{Y,k,nv} \\ & & & & \mathbf{0}\end{pmatrix}}.$$

The block $\mathbf{D}_{i,j}^{Y,k,nv}$ contains the transition, when there are *k* units in the system, from *i* units in the repair facility to *j* (a type event *Y* occurs) and the repairperson is in his workplace. For instance the cases $\mathbf{D}_{01}^{A,k,nv}$ and $\mathbf{D}_{01}^{B,k,nv}$ (transition $E_{0}^{k,nv}\rightarrow E_{1}^{k,nv}$ for type *A* and *B* respectively) are analyzed.

In both cases, there are *k* units in the system and none of them is in the repair facility (all operational). The online unit goes to the repair facility if it undergoes an internal repairable failure ($\mathbf{H}_{A}$) or a major inspection ($\mathbf{H}_{B}$). In both cases a new unit will occupy the online place if the number of units in the system is greater than one. If the event is a repairable failure, then the unit will begin the repair given that the repairperson is not on vacations ($\boldsymbol{\beta}_{1}$). If the event was major inspection, the initial distribution for the preventive maintenance would be $\boldsymbol{\beta}_{2}$.

- If the number of units is greater or equal than *R*, the repairperson can be on vacation or not. If the repairperson returns and there are less than *R* operational units then he remains at his workplace. Given that these events *A* and *B* occurs when a repairable or major inspection occurs (without returning to work) then, for *k* = *R*,…, *n* ( $N=k-R+1$, limit of number of units in the repair facility to remain the repairperson):

$$
\mathbf{D}_k^Y = \begin{array}{c} \\ E_0^{k,v} \\ E_1^{k,v} \\ \vdots \\ E_{N-1}^{k,v} \\ E_N^{k,v} \\ E_{N+1}^{k,v} \\ \vdots \\ E_{k-1}^{k,v} \\ E_k^{k,v} \\ E_N^{k,nv} \\ E_{N+1}^{k,nv} \\ \vdots \\ E_{k-1}^{k,nv} \\ E_k^{k,nv} \end{array}
\begin{array}{c}
\begin{array}{cccccccccc|ccccc}
E_0^{k,v} & E_1^{k,v} & \cdots & E_{N-1}^{k,v} & E_N^{k,v} & E_{N+1}^{k,v} & \cdots & E_{k-1}^{k,v} & E_k^{k,v} & & E_N^{k,nv} & E_{N+1}^{k,nv} & E_{N+2}^{k,nv} & E_{k-1}^{k,nv} & E_k^{k,nv}
\end{array} \\
\left(\begin{array}{ccccccccc|ccccc}
\mathbf{0} & \mathbf{D}_{0,1}^{Y,k,v} & \cdots & & & & & & & & & & & \\
 & \mathbf{0} & \ddots & & & & & & & & & & & \\
 & & \ddots & \ddots & & & & & & & & & & \\
 & & & \mathbf{0} & \mathbf{D}_{N-1,N}^{Y,k,v} & & & & & & & & & \\
 & & & & \mathbf{0} & \mathbf{D}_{N,N+1}^{Y,k,v} & & & & & & & & \\
 & & & & & \mathbf{0} & \ddots & & & & & & & \\
 & & & & & & \ddots & \ddots & & & & & & \\
 & & & & & & & \mathbf{0} & \mathbf{D}_{k-1,k}^{Y,k,v} & & & & & \\
 & & & & & & & & \mathbf{0} & & & & & \\
\hline
 & & & & & & & & & \mathbf{D}_{N,N}^{Y,k,nv} & \mathbf{D}_{N,N+1}^{Y,k,nv} & & & \\
 & & & & & & & & & & \mathbf{D}_{N+1,N+1}^{Y,k,nv} & \mathbf{D}_{N+1,N+2}^{Y,k,nv} & & \\
 & & & & & & & & & & & \ddots & \ddots & \\
 & & & & & & & & & & & & \mathbf{D}_{k-1,k-1}^{Y,k,nv} & \mathbf{D}_{k-1,k}^{Y,k,nv} \\
 & & & & & & & & & & & & & \mathbf{0}
\end{array}\right)
\end{array}
$$

This matrix is partitioned into two great matrix blocks depending on the transition between macro states; continues on vacation and continues on the repair facility.

The block $\mathbf{D}_{i,j}^{Y,k,v}$ contains the transition, when there are $k$ units in the system, from $i$ units in the repair facility to $j$ (type $Y$) and the repairperson continues on vacation. For instance, the cases $\mathbf{D}_{01}^{A,k,v}$ and $\mathbf{D}_{01}^{B,k,v}$ corresponds to the transition $E_0^{k,v} \rightarrow E_1^{k,v}$ for type $A$ and $B$ respectively.

These matrices are for $k = 1,\ldots, n$ and $R > 1$

$$\mathbf{D}_{01}^{A,k,nv} = \left(\left(I_{\{k>1\}}\mathbf{H}_A + I_{\{k=1\}}\mathbf{H}'_A\right)\otimes\boldsymbol{\beta}_1, \mathbf{0}\right) \quad ; \quad \mathbf{D}_{01}^{B,k,nv} = \left(\mathbf{0}, \left(I_{\{k>1\}}\mathbf{H}_B + I_{\{k=1\}}\mathbf{H}'_B\right)\otimes\boldsymbol{\beta}_2\right).$$

The rest of matrices for this matrix block are the following.

$$\mathbf{D}_{1,1}^{A,k,nv} = \begin{pmatrix} \mathbf{H}_A \otimes \mathbf{S}_1^0 \otimes \boldsymbol{\beta}_1 & \mathbf{0} \\ \mathbf{H}_A \otimes \mathbf{S}_2^0 \otimes \boldsymbol{\beta}_1 & \mathbf{0} \end{pmatrix} \quad ; \quad \mathbf{D}_{1,1}^{B,k,nv} = \begin{pmatrix} \mathbf{0} & \mathbf{H}_B \otimes \mathbf{S}_1^0 \otimes \boldsymbol{\beta}_2 \\ \mathbf{0} & \mathbf{H}_B \otimes \mathbf{S}_2^0 \otimes \boldsymbol{\beta}_2 \end{pmatrix}$$

For $r = 2, \ldots, \ k-1$

$$\mathbf{D}_{r,r}^{A,k,nv} = \begin{pmatrix} I_{2^{r-2}} \otimes \left(\mathbf{H}_A \otimes \mathbf{S}_1^0 \otimes \boldsymbol{\beta}_1, \mathbf{0}\right) & \mathbf{0} \\ \mathbf{0} & I_{2^{r-2}} \otimes \left(\mathbf{H}_A \otimes \mathbf{S}_1^0 \otimes \boldsymbol{\beta}_2, \mathbf{0}\right) \\ I_{2^{r-2}} \otimes \left(\mathbf{H}_A \otimes \mathbf{S}_2^0 \otimes \boldsymbol{\beta}_1, \mathbf{0}\right) & \mathbf{0} \\ \mathbf{0} & I_{2^{r-2}} \otimes \left(\mathbf{H}_A \otimes \mathbf{S}_2^0 \otimes \boldsymbol{\beta}_2, \mathbf{0}\right) \end{pmatrix}$$

$$\mathbf{D}_{r,r}^{B,k,nv} = \begin{pmatrix} I_{2^{r-2}} \otimes \left(\mathbf{0}, \mathbf{H}_B \otimes \mathbf{S}_1^0 \otimes \boldsymbol{\beta}_1\right) & \mathbf{0} \\ \mathbf{0} & I_{2^{r-2}} \otimes \left(\mathbf{0}, \mathbf{H}_B \otimes \mathbf{S}_1^0 \otimes \boldsymbol{\beta}_2\right) \\ I_{2^{r-2}} \otimes \left(\mathbf{0}, \mathbf{H}_B \otimes \mathbf{S}_2^0 \otimes \boldsymbol{\beta}_1\right) & \mathbf{0} \\ \mathbf{0} & I_{2^{r-2}} \otimes \left(\mathbf{0}, \mathbf{H}_B \otimes \mathbf{S}_2^0 \otimes \boldsymbol{\beta}_2\right) \end{pmatrix}$$

For $r = \max\{1, k-R+1\}, \ldots, \ k-1$

$$\mathbf{D}_{r,r+1}^{A,k,nv} = \begin{pmatrix} \mathbf{I}_{2^{r-1}} \otimes \left(\left(I_{\{r<k-1\}}\mathbf{H}_A + I_{\{r=k-1\}}\mathbf{H}'_A\right)\otimes\mathbf{S}_1, \mathbf{0}\right) & \mathbf{0} \\ \mathbf{0} & \mathbf{I}_{2^{r-1}} \otimes \left(\left(I_{\{r<k-1\}}\mathbf{H}_A + I_{\{r=k-1\}}\mathbf{H}'_A\right)\otimes\mathbf{S}_2, \mathbf{0}\right) \end{pmatrix}$$

$$\mathbf{D}_{r,r+1}^{B,k,nv} = \begin{pmatrix} \mathbf{I}_{2^{r-1}} \otimes \left(\mathbf{0}, \left(I_{\{r<k-1\}}\mathbf{H}_B + I_{\{r=k-1\}}\mathbf{H}'_B\right)\otimes\mathbf{S}_1\right) & \mathbf{0} \\ \mathbf{0} & \mathbf{I}_{2^{r-1}} \otimes \left(\mathbf{0}, \left(I_{\{r<k-1\}}\mathbf{H}_A + I_{\{r=k-1\}}\mathbf{H}'_A\right)\otimes\mathbf{S}_2\right) \end{pmatrix}$$

For $r = 1,\ldots, k-1$ and $k \geq R$

$$\mathbf{D}_{0,1}^{A,k,v} = \left(\mathbf{H}_A \otimes \left(\mathbf{V} + I_{\{k\geq R+1\}}\mathbf{V}^0\mathbf{v}\right), \mathbf{0}\right); \quad \mathbf{D}_{0,1}^{B,k,v} = \left(\mathbf{0}, \mathbf{H}_B \otimes \left(\mathbf{V} + I_{\{k\geq R+1\}}\mathbf{V}^0\mathbf{v}\right)\right)$$

$$\mathbf{D}_{r,r+1}^{A,k,v} = \mathbf{I}_{2^r} \otimes \left( \left( I_{\{r<k-1\}} \mathbf{H}_A + I_{\{r=k-1\}} \mathbf{H}'_A \right) \otimes \left( \mathbf{V} + I_{\{r<N-1\}} \mathbf{V}^0 \mathbf{v} \right), \mathbf{0} \right)$$

$$\mathbf{D}_{r,r+1}^{B,k,v} = \mathbf{I}_{2^r} \otimes \left( \mathbf{0}, \left( I_{\{r<k-1\}} \mathbf{H}_B + I_{\{r=k-1\}} \mathbf{H}'_B \right) \otimes \left( \mathbf{V} + I_{\{r<N-1\}} \mathbf{V}^0 \mathbf{v} \right) \right).$$

## 4. Measures

Multiple interesting measures in transient and stationary regime can be worked out and are described in this section.

### *4.1. The transient and the stationary distribution*

The transient distribution is determined by the initial distribution and the transition probability matrix of the vector Markov process given in Section 3.3.

Initially the online unit is new and the inspection time begins. Then, the initial distribution of the Markov process is $\phi = [\boldsymbol{\alpha} \otimes \boldsymbol{\gamma}_{st} \otimes \boldsymbol{\eta}, \mathbf{0}]$ where $\boldsymbol{\gamma}_{st}$ is the stationary distribution of the phase-type renewal process with transition probability matrix $\mathbf{L} + \mathbf{L}^0 \boldsymbol{\gamma}$. Therefore, $\boldsymbol{\gamma}_{st} = [1, \mathbf{0}] \left[ \mathbf{e} \,|\, \left( \mathbf{L} + \mathbf{L}^0 \boldsymbol{\gamma} - \mathbf{I} \right)^* \right]^{-1}$ .

The probability of occupying the macro-state $E_s^{k,a}$ at time $v$ is worked out by matrix blocks as $\mathbf{p}_{E_s^{k,a}}^{v} = \left( \phi \mathbf{D}^{v} \right)_{I_s^{k,a}}$ where $I_s^{k,a}$ indicates the range for the corresponding states. Evidently, $\mathbf{p}^{v}$ is the transient distribution at time $v$.

To calculate the stationary distribution in a matrix-algorithmic form, we have partitioned the matrix $\mathbf{D}$ for the transitions between the macro-states $\mathbf{U}^j$ into the following blocks,

$$\mathbf{D} = \begin{pmatrix} \mathbf{D}_{n,n} & \mathbf{D}_{n,n-1} & \mathbf{0} & \cdots & \mathbf{0} & \mathbf{0} \\ \mathbf{0} & \mathbf{D}_{n-1,n-1} & \mathbf{D}_{n-1,n-2} & \cdots & \mathbf{0} & \mathbf{0} \\ \vdots & \vdots & \ddots & \ddots & \vdots & \vdots \\ \mathbf{0} & \mathbf{0} & \cdots & \cdots & \mathbf{D}_{22} & \mathbf{D}_{21} \\ \mathbf{D}_{1n} & \mathbf{0} & \cdots & \cdots & \cdots & \mathbf{D}_{11} \end{pmatrix},$$

where

$$\mathbf{D}_{ii} = \mathbf{D}_i^{O} + \mathbf{D}_i^{A} + \mathbf{D}_i^{B} + \mathbf{D}_i^{D} + \mathbf{D}_i^{AD} + \mathbf{D}_i^{BD} \quad ; \quad i = 1, \ldots, n$$

$$\mathbf{D}_{i,i-1} = \mathbf{D}_i^{C} + \mathbf{D}_i^{CD} \quad ; \quad i = 2, \ldots, n$$

$$\mathbf{D}_{1,n} = \mathbf{D}_1^{NS}.$$

The stationary distribution $\boldsymbol{\pi}$ verifies the balance equations $\boldsymbol{\pi}\mathbf{D} = \boldsymbol{\pi}$ and the normalization equation $\boldsymbol{\pi}\mathbf{e} = 1$. This vector is partitioned into the macro-states $\mathbf{U}^j$, $j$ units in the system, then, $\boldsymbol{\pi} = \{ \boldsymbol{\pi}_n, \boldsymbol{\pi}_{n-1}, \ldots, \boldsymbol{\pi}_1 \}$ for the macro-states $\mathbf{U}^n$, …, $\mathbf{U}^1$, respectively.

The solution of this matrix system is

$$\boldsymbol{\pi}_j = \boldsymbol{\pi}_1 \mathbf{R}_j \quad ; \quad j = 2, \ldots, n,$$

being

$$\mathbf{R}_j = \mathbf{R}_{j+1} \mathbf{G}_{j+1,j} = \mathbf{G}_{1n} \mathbf{G}_{n,n-1} \cdots \mathbf{G}_{j+1,j} \quad ; \quad j = 2, \ldots, n-1$$

$$\mathbf{R}_n = \mathbf{G}_{1,n}$$

and

$$\mathbf{G}_{ij} = \mathbf{D}_{ij} \left( \mathbf{I} - \mathbf{D}_{jj} \right)^{-1} \text{ for } (i,j) \in \{ (1,n), (n,n-1), (n-1,n-2), \ldots, (3,2) \} .$$

The transition probability vector for the macro-state $\mathbf{U}^1$ can be worked out from the normalization condition and one balance equation as

$$\boldsymbol{\pi}_1 = (1, \mathbf{0}) \left( \mathbf{e} + \sum_{j=2}^{n} \mathbf{R}_j \mathbf{e} \,\middle|\, \left( \mathbf{I} - \mathbf{D}_{11} - \mathbf{R}_2 \mathbf{D}_{21} \right)^* \right)^{-1},$$

where * is the corresponding matrix without the first column.

From the stationary distribution and considering the macro-states, multiple proportional time measures can be defined:

- Proportional time that the system has $k$ units: $\boldsymbol{\pi}_{\mathbf{U}^k}$ .
- Proportional time that the repairperson is in the workplace:

$$\Upsilon_{nv} = \sum_{k=1}^{R-1}\sum_{s=0}^{k}\boldsymbol{\pi}_{E_s^{k,nv}}\mathbf{e} + \sum_{k=R}^{n}\sum_{s=k-R+1}^{k}\boldsymbol{\pi}_{E_s^{k,nv}}\mathbf{e}\,.$$

- Proportional time that the repairperson is on vacation:
  $\Upsilon_v = 1 - \Upsilon_{nv}$
- Proportional time that the repairperson is working:
  $\Upsilon_w = \sum_{k=1}^{R-1}\sum_{s=1}^{k}\boldsymbol{\pi}_{E_s^{k,nv}}\mathbf{e} + \sum_{k=R}^{n}\sum_{s=k-R+1}^{k}\boldsymbol{\pi}_{E_s^{k,nv}}\mathbf{e}$
- Proportional time that the repairperson is idle:
  $\Upsilon_i = \Upsilon_{nv} - \Upsilon_w$.

### *4.2. Availability and mean times*

It is very interesting to calculate the availability of the system, the mean time in each macro-state and the mean operational time. It has been summed up in Table 1 in both regimes, transient and stationary.

**Table 1**. Availability and mean times in transient and stationary regime

| | **Transient regime (up to time ν)** | **Stationary regime** |
|---|---|---|
| **Availability** | $A(\nu) = 1 - \sum_{k=R}^{n}\left(\mathbf{p}^{\nu}_{E_k^{k,\nu}}\cdot\mathbf{e} + \mathbf{p}^{\nu}_{E_k^{k,nv}}\cdot\mathbf{e}\right) - \sum_{k=1}^{R-1}\mathbf{p}^{\nu}_{E_k^{k,nv}}\cdot\mathbf{e}$ | $A = 1 - \sum_{k=R}^{n}\left(\boldsymbol{\pi}_{E_k^{k,\nu}}\cdot\mathbf{e} + \boldsymbol{\pi}_{E_k^{k,nv}}\cdot\mathbf{e}\right) - \sum_{k=1}^{R-1}\boldsymbol{\pi}_{E_k^{k,nv}}\cdot\mathbf{e}$ |
| **Mean time in $\mathbf{E}_s^{k,\nu};\mathbf{E}_s^{k,nv}$** | $\psi_{k,s}(\nu) = \sum_{m=0}^{\nu}\left(\mathbf{p}^{m}_{E_s^{k,\nu}}\cdot\mathbf{e} + \mathbf{p}^{m}_{E_s^{k,nv}}\cdot\mathbf{e}\right)$ | $\psi_{k,s} = \boldsymbol{\pi}_{E_s^{k,\nu}}\cdot\mathbf{e} + \boldsymbol{\pi}_{E_s^{k,nv}}\cdot\mathbf{e}$ |
| **Mean time in $\mathbf{E}^k$** | $\psi_k(\nu) = \sum_{s=0}^{k}\psi_{k,s}(\nu)$ | $\psi_k = \sum_{s=0}^{k}\psi_{k,s}$ |
| **Mean operational time** | $\mu_{op}(\nu) = \sum_{k=1}^{K}\sum_{s=0}^{k-1}\psi_{k,s}(\nu)$ | $\mu_{op} = \sum_{k=1}^{K}\sum_{s=0}^{k-1}\psi_{k,s}$ |

### *4.3. Time up to first time that the system is replaced*

A system composed of $n$ units is replaced by a new and identical one when all units undergo a non-repairable failure. The time up to this event is phase-type distributed with representation $(\phi, \mathbf{D}')$ where $\mathbf{D}' = \mathbf{D}^O + \mathbf{D}^A + \mathbf{D}^B + \mathbf{D}^C + \mathbf{D}^D + \mathbf{D}^{AD} + \mathbf{D}^{BD} + \mathbf{D}^{CD}$.

### *4.4. Expected number of events*

The expected number of events up to time $\nu$ is determined using the Markovian Arrival Process with Marked arrivals developed in Section 3.3. If the event considered is denoted by $Y$ then the corresponding expected number of events is given by

$$\Lambda^Y(\nu) = \sum_{u=1}^{\nu}\mathbf{p}^{u-1}\mathbf{D}^Y\mathbf{e},$$

for $Y = A, B, C, D, AD, BD, CD, NS$. This value in stationary regime is $\Lambda^Y = \boldsymbol{\pi}\mathbf{D}^Y\mathbf{e}$.

Another mean number of events can be calculated as follows.

*Mean number of repairable failures*

A repairable failure can occur when the repairperson resumes to work or not at the same time. Then, the mean number up to time $\nu$ is

$\Lambda^{rep}(\nu) = \sum_{u=1}^{\nu} \mathbf{p}^{u-1}\left(\mathbf{D}^{A} + \mathbf{D}^{AD}\right)\mathbf{e}$ and in stationary regime it is $\Lambda^{rep} = \boldsymbol{\pi}\ \mathbf{D}^{A} + \mathbf{D}^{AD}\ \mathbf{e}$.

*Mean number of major inspections*

Analogously to the repairable case, a major inspection can occur when the repairperson occupies the workplace or not at the same time. Then, it is in transient regime

$\Lambda^{mi}(\nu) = \sum_{u=1}^{\nu} \mathbf{p}^{u-1}\left(\mathbf{D}^{B} + \mathbf{D}^{BD}\right)\mathbf{e}$ and in the stationary case it is $\Lambda^{mi} = \boldsymbol{\pi}\ \mathbf{D}^{B} + \mathbf{D}^{BD}\ \mathbf{e}$.

*Mean number of non-repairable failures (no provoking system failure)*

The mean number of non-repairable failures up to time $\nu$ is

$$\Lambda^{nr}(\nu) = \sum_{u=1}^{\nu} \mathbf{p}^{u-1}\left(\mathbf{D}^{C} + \mathbf{D}^{CD}\right)\mathbf{e}.$$

This value in the stationary case is $\Lambda^{nr} = \boldsymbol{\pi}\ \mathbf{D}^{C} + \mathbf{D}^{CD}\ \mathbf{e}$.

*Mean number of times that the repairperson resumes to work*

The mean number that the repairperson resumes and remains in his workplace up to a certain time is given by

$$\Lambda^{rejoined}(\nu) = \sum_{u=1}^{\nu} \mathbf{p}^{u-1}\left(\mathbf{D}^{D} + \mathbf{D}^{AD} + \mathbf{D}^{BD} + \mathbf{D}^{CD}\right)\mathbf{e}.$$

In the stationary case it is $\Lambda^{rejoined} = \boldsymbol{\pi}\ \mathbf{D}^{D} + \mathbf{D}^{AD} + \mathbf{D}^{BD} + \mathbf{D}^{CD}\ \mathbf{e}$.

*Mean number of times that the repairperson resumes and begins a new period of vacation*

The mean number that the repairperson resumes and begins a new period of vacation up to a certain time is given by

$$\Lambda^{r-b}(\nu) = \sum_{u=1}^{\nu} \mathbf{p}^{u-1}\mathbf{Q}\mathbf{e}.$$

where **Q** is a matrix described in Appendix C. In stationary case it is $\Lambda^{r-b} = \boldsymbol{\pi}\mathbf{Q}\mathbf{e}$.

*Mean number of new systems*

When the system is composed of only one unit and a non-repairable failure occurs, the system is restarted with *n* new units. The mean number of new systems up to time $\nu$ is

$$\Lambda^{NS}(\nu) = \sum_{u=1}^{\nu} \mathbf{p}^{u-1}\mathbf{D}^{NS}\mathbf{e}.$$

This measure in stationary case is $\Lambda^{NS} = \boldsymbol{\pi}\mathbf{D}^{NS}\mathbf{e}$.

## 5. Rewards and Costs

To analyze the effectiveness of the model from an economic point of view, costs and rewards have been taken into account. A net profit vector associated to the state-space is built. Previously, multiple values are introduced:

*B*: Gross profit per unit of time if the system is operational.

$\mathbf{c}_0$: expected cost per unit of time depending on the operational phase while the system is operational.

$\mathbf{cr}_1$: expected corrective repair cost per unit of time depending on the repair phase.

$\mathbf{cr}_2$: expected preventive maintenance cost per unit of time for a unit that was observed with major damage depending on the preventive maintenance phase.

*H*: repairperson cost per unit of time while the repairperson in idle.

*C*: loss per unit of time while the system is not operational

*G*: fixed cost associated to each return of the repairperson (independently of if he stays or not).

*fcr*: fixed cost each time that the online unit undergoes a repairable failure from the online unit

*fmi*: fixed cost each time that the online unit undergoes a major inspection

*fnu*: cost for a new unit (*n*·*fnu* cost of a new system)

*5.1. Net profit vector*

When the system occupies a determined state, a net profit value is produced. Costs and rewards from the online unit and the cost provoked by the repairperson have been taken into account to build the net profit vector.

*Online unit*

If only the online unit is considered when the system visits the macro-state $\mathbf{E}_s^{k,nv}$, a net reward for the phases of this macro-state is worked out. The profit net vector for the online unit if the repairperson is on his workplace ($\mathbf{E}_s^{k,nv}$) is for $k = 1,\ldots, n$,

$$\mathbf{nr}_s^{k,nv} = \begin{cases} B\mathbf{e}_{mt\varepsilon} - \mathbf{c_0} \otimes \mathbf{e}_{t\varepsilon} & ; \quad s = 0 \\ B\mathbf{e}_{mt\varepsilon 2^{s-1}(z_1+z_2)} - \mathbf{c_0} \otimes \mathbf{e}_{t\varepsilon 2^{s-1}(z_1+z_2)} & ; \quad s = 1,\ldots,k-1 \\ -C \cdot \mathbf{e}_{t2^{s-1}(z_1+z_2)} & ; \quad s = k. \end{cases}$$

It can be expressed for any number of units in the repair facility as the following column vector $\mathbf{nr}_{Total}^{k,nv} = \left(\mathbf{nr}_0^{k,nv}{}';\ldots;\mathbf{nr}_k^{k,nv}{}'\right)'$.

If the number of units in the repair facility is $N$ or more, then the repairperson remains at his workplace without vacation. In this case we define $\mathbf{nr}_{fromN}^{k,nv} = \left(\mathbf{nr}_N^{k,nv}{}';\ldots;\mathbf{nr}_k^{k,nv}{}'\right)'$.

For the case when the repairperson is on vacation, the profit net vector for the online unit for the macro-state $\mathbf{E}_s^{k,v}$ is

$$\mathbf{nr}_s^{k,v} = \begin{cases} B\mathbf{e}_{mt\varepsilon\upsilon} - \mathbf{c_0} \otimes \mathbf{e}_{t\varepsilon\upsilon} & ; \quad s = 0 \\ B\mathbf{e}_{mt\varepsilon\upsilon 2^s} - \mathbf{c_0} \otimes \mathbf{e}_{t\varepsilon\upsilon 2^s} & ; \quad s = 1,\ldots,k-1 \\ -C \cdot \mathbf{e}_{t\upsilon 2^s} & ; \quad s = k. \end{cases}$$

For any number of units in the repair facility the column vector $\mathbf{nr}_{Total}^{k,v} = \left(\mathbf{nr}_0^{k,v}{}';\ldots;\mathbf{nr}_k^{k,v}{}'\right)'$ is defined.

Then, if the total state space is considered then the net reward, according to the state visited, for the online unit is $\mathbf{nr} = \left(\mathbf{nr}_{Total}^{n,v}{}';\mathbf{nr}_{fromN}^{n,nv}{}';\mathbf{nr}_{Total}^{n-1,v}{}';\mathbf{nr}_{fromN}^{n-1,nv}{}';\ldots;\mathbf{nr}_{Total}^{N,v}{}';\mathbf{nr}_{fromN}^{N,nv}{}';\mathbf{nr}_{Total}^{N-1,nv}{}';\mathbf{nr}_{Total}^{N-1,nv}{}';\ldots;\mathbf{nr}_{Total}^{1,nv}{}'\right)'$.

*Repair facility*

If only the repair facility is considered, when the system visits the macro-states $\mathbf{E}_s^{k,nv}$, a cost vector for the phases of the corresponding macro-state, for $k = 1,\ldots, n$ is

$$\mathbf{nc}_s^{k,nv} = \begin{cases} H \cdot \mathbf{e}_{mt\varepsilon} & ; \quad s = 0 \\ \mathbf{e}_{t(m\varepsilon)^{I_{\{s<k\}}}} \otimes \begin{pmatrix} \mathbf{e}_{2^{s-1}} \otimes \mathbf{cr}_1 \\ \mathbf{e}_{2^{s-1}} \otimes \mathbf{cr}_2 \end{pmatrix} & ; \quad s = 1,\ldots,k. \end{cases}$$

For any number of units in the repair facility, the following column vectors are defined,
$\mathbf{nc}_{Total}^{k,nv} = \left(\mathbf{nc}_0^{k,nv}{}';\ldots;\mathbf{nc}_k^{k,nv}{}'\right)'$, $\mathbf{nc}_{fromN}^{k,nv} = \left(\mathbf{nc}_N^{k,nv}{}';\ldots;\mathbf{nc}_k^{k,nv}{}'\right)'$.

For any $k$ and $s$ while the repairperson is on vacation the cost of the repair facility is zero, then the following column vector is defined for this case $\mathbf{nc}_s^{k,v} = \mathbf{0}_{(m\varepsilon)^{I_{\{s<k\}}}t\upsilon 2^s}$. For any number of units in the repair facility it is defined $\mathbf{nc}_{Total}^{k,v} = \left(\mathbf{nc}_0^{k,v}{}';\ldots;\mathbf{nc}_k^{k,v}{}'\right)'$.

Then, the cost vector associated to the state space due to repair is given by

$$\mathbf{nc} = \left(\mathbf{nc}^{n,v}{}';\mathbf{nc}_{fromN}^{n,nv}{}';\mathbf{nc}^{n-1,v}{}';\mathbf{nc}_{fromN}^{n-1,nv}{}';\ldots;\mathbf{nc}^{N,v}{}';\mathbf{nc}_{fromN}^{N,nv}{}';\mathbf{nc}_{Total}^{N-1,nv}{}';\mathbf{nc}_{Total}^{N-1,nv}{}';\ldots;\mathbf{nc}_{Total}^{1,nv}{}'\right)'.$$

Therefore, the net profit vector corresponding to the online unit and the repair facility for the global state space is given by

$$\mathbf{c} = \mathbf{nr} - \mathbf{nc} = \begin{pmatrix} \mathbf{c}^n \\ \mathbf{c}^{n-1} \\ \vdots \\ \mathbf{c}^1 \end{pmatrix},$$

where
$\mathbf{c}^k = \left(\mathbf{nr}_{Total}^{k,nv}{}' - \mathbf{nc}_{Total}^{k,nv}{}'\right)'$ for $k = 1,\ldots, R-1$,
$\mathbf{c}^k = \left(\mathbf{nr}^{k,v}{}' - \mathbf{nc}^{k,v}{}' ; \mathbf{nc}_{fromN}^{k,nv}{}' - \mathbf{nc}_{fromN}^{k,nv}{}'\right)'$ for $k = R,\ldots, n$.

*5.2. Expected net profits and total net profit*

Net reward measures are worked out, in transient and stationary regime, to analyze the effectiveness of the system from an economic point of view.

*Expected net profit from the online unit up to time $\nu$*

The expected net profit up to time ν by considering only the online unit is

$$\Phi_w^{\nu} = \sum_{m=0}^{\nu} \mathbf{p}^m \cdot \mathbf{nr}.$$

In stationary regime it is given by $\Phi_{w_s} = \boldsymbol{\pi} \cdot \mathbf{nr}$.

*Expected cost from corrective repair and preventive maintenance*

The expected cost because of corrective repair and preventive maintenance up to time $\nu$ is calculated. It is respectively

$$\Phi_{cr}^{\nu} = \sum_{m=0}^{\nu} \mathbf{p}^m \cdot \mathbf{mc}^{cr} \quad \text{and} \quad \Phi_{pm}^{\nu} = \sum_{m=0}^{\nu} \mathbf{p}^m \cdot \mathbf{mc}^{pm} \text{ where}$$

where $\mathbf{mc}^{cr}$ is the vector $\mathbf{nc}$ with $\mathbf{cr}_2 = \mathbf{0}_{z_2}$ and $\mathbf{mc}^{pm}$ is the vector $\mathbf{nc}$ with $\mathbf{cr}_1 = \mathbf{0}_{z_1}$, being $\mathbf{0}_a$ a column vector of 0's with order *a*.

If the stationary regime is considered, then

$$\Phi_{cr_s} = \boldsymbol{\pi} \cdot \mathbf{mc}^{cr} \quad \text{and} \quad \Phi_{pm_s} = \boldsymbol{\pi} \cdot \mathbf{mc}^{pm} \quad .$$

*Total net profit*

If costs, fixed costs and profits are considered, the total net profit up to time $\nu$ is

$$\Phi^{\nu} = \Phi_w^{\nu} - \Phi_{cr}^{\nu} - \Phi_{pm}^{\nu} - \left(1 + \Lambda^{NS}(\nu)\right) \cdot n \cdot fnu - \Lambda^{rep}(\nu) \cdot fcr - \Lambda^{mi}(\nu) \cdot fmi - \Lambda^{r-b}(\nu) \cdot G$$

In the stationary case it is

$$\Phi = \Phi_w - \Phi_{cr} - \Phi_{pm} - \left(1 + \Lambda^{NS}\right) \cdot n \cdot fnu - \Lambda^{rep} \cdot fcr - \Lambda^{mi} \cdot fmi - \Lambda^{r-b} \cdot G.$$

## 6. A numerical example

The system modelled in this paper can be applied to real-world engineering problems. It would be interesting to examine whether or not preventive maintenance is profitable and to determine the optimum distribution for vacation time and hence the corresponding value of *R*.

*6.1. The system*

We assume a standby system composed of 4 units initially as described in this work. Each unit is composed of 4 performance internal states where the first two are considered minor damage and the last two, major damage. The transition probability matrix for wearing out time is given by

$$\mathbf{T} = \begin{pmatrix} 0.96 & 0.03 & 0 & 0 \\ 0 & 0.97 & 0.01 & 0 \\ 0 & 0 & 0.85 & 0.06 \\ 0 & 0 & 0 & 0.6 \end{pmatrix},$$

beginning in the initial state ($\alpha$=(1,0,0,0)). From each state, only a transition to failure or to next performance level state can occur. The transition probability to repairable and non-repairable failure depending on the performance state are given by the column vectors

$$\mathbf{T}_r^0 = \begin{pmatrix} 0.008 \\ 0.016 \\ 0.072 \\ 0.32 \end{pmatrix} \text{ and } \mathbf{T}_{nr}^0 = \begin{pmatrix} 0.002 \\ 0.004 \\ 0.018 \\ 0.080 \end{pmatrix} \text{ respectively.}$$

The online unit is subject to external shocks. The time between two consecutive external shocks follows a phase-type distribution with representation ($\boldsymbol{\gamma}$, $\mathbf{L}$) being

$$\boldsymbol{\gamma} = (1,0) \text{ and } \mathbf{L} = \begin{pmatrix} 0.9 & 0.05 \\ 0 & 0.5 \end{pmatrix}.$$

The mean time between two consecutive accidental external failures is equal to 11 units of time.

Each time that the system suffers an external shock the internal performance can be modified by producing even a repairable or non-repairable failure. The matrix that governs the changes into the operational states is

$$\mathbf{W} = \begin{pmatrix} 0.2 & 0.1 & 0.3 & 0.1 \\ 0 & 0.1 & 0.3 & 0.1 \\ 0 & 0 & 0.3 & 0.1 \\ 0 & 0 & 0 & 0.1 \end{pmatrix},$$

and the change to a repairable and non-repairable is

$$\mathbf{W}_r^0 = \begin{pmatrix} 0.3 \\ 0.4 \\ 0.5 \\ 0.6 \end{pmatrix} \text{ and } \mathbf{W}_{nr}^0 = \begin{pmatrix} 0 \\ 0.1 \\ 0.1 \\ 0.3 \end{pmatrix} \text{ respectively.}$$

Also, when an external shock occurs, a total failure can be produced with a probability equal to $\omega^0 = 0.2$.

Inspections occur randomly where the inter-inspection time is phase-type distributed with representation $(\boldsymbol{\eta}, \mathbf{M})$ being

$$\boldsymbol{\eta} = (1,0), \ \mathbf{M} = \begin{pmatrix} 0.85 & 0.1 \\ 0.45 & 0.4 \end{pmatrix}.$$

When a unit undergoes a repairable failure or inspection observes major damage, this one goes to the repair facility. Therefore, two types of tasks can be developed by the repairperson, corrective repair and preventive maintenance. Both are phase-type distributed with representation for the corrective repair time,

$$\boldsymbol{\beta}_1 = (1,0,0) \text{ and } \mathbf{S}_1 = \begin{pmatrix} 0.2 & 0.4 & 0.3 \\ 0.2 & 0.2 & 0.5 \\ 0.3 & 0.2 & 0.3 \end{pmatrix}$$

and for the preventive maintenance time,

$$\boldsymbol{\beta}_2 = (1,0,0) \text{ and } \mathbf{S}_2 = \begin{pmatrix} 0.2 & 0.3 & 0.1 \\ 0.1 & 0.1 & 0.4 \\ 0.2 & 0.2 & 0.2 \end{pmatrix}.$$

The mean corrective repair time is 7.3810 units of time and for the preventive maintenance case it is equal to 2.5 units of time.

### *6.2. Costs and rewards*

Different costs and rewards have been considered as described in Section 5. We assume a gross profit while the system is operational equal to $B$=60. This is also the loss per unit of time while the system is not operational, $C$=60. The online unit has a cost while is operational depending on the operational phase. This vector is $\mathbf{c}_0 = (5,12,30,40)'$. The repairperson can be on vacation or on his workplace. Each time that the repairperson returns on his vacation a cost equal to $G$=20 is produced. While the repairperson is idle, a cost equal to $H$=15 is produced.

The online unit can undergo a repairable failure. In this case, the unit goes to the repair facility for corrective repair. A fix cost is considered for each failure equal to $fcr$=10. Once in corrective repair, a cost depending on the state is given by $\mathbf{cr}_1 = (18,18,18)'$.

Also, when inspection observes major damage, the unit goes to the repair facility for preventive maintenance. A fixed cost is produced, *fmi* = 5. Once in the repair facility the cost will depend on the preventive maintenance state. It is given by the vector $\mathbf{cr}_2 = (15.5, 15.5, 15.5)'$. Finally, when all units undergo a non-repairable failure the system is re-started. It has a cost per unit equal to *fnu* = 100.

### 6.3. Optimization analysis

The repairperson can take a vacation, for a random duration, and inspections may take place at random intervals. This circumstance raises two interesting questions. Firstly, if a distribution class is assumed for the duration of the vacation, from an economic standpoint what is the optimum distribution and the optimum value of $R$ (i.e. the limit value of the number of operational units needed to require the repairperson to remain in the facility on returning from vacation) from an economic standpoint? Secondly, is it profitable to perform preventive maintenance?

To answer these questions, we consider two classes of distributions, the geometric distribution and the Erlang distribution, from which optimum values for $R$ and the other parameters can be determined.

*The Geometric distribution case*

We assume that the vacation time of the repairperson is distributed geometrically with parameter $p$. Then, the p.m.f. is

$$P\{X = n\} = p^{n-1}(1-p) \quad ; n = 0, 1, 2, \ldots$$

The stationary net profit depending on $p$ for the system with and without preventive maintenance is shown in Figure 4. It has been worked out from Section 5.2. We can see that, when the geometric distribution is considered, the optimum value is reached for the preventive maintenance case with $p$ = 0.8 and $R$=3. In this case, in stationary case, the net profit per unit of time would be equal to 22.0571.

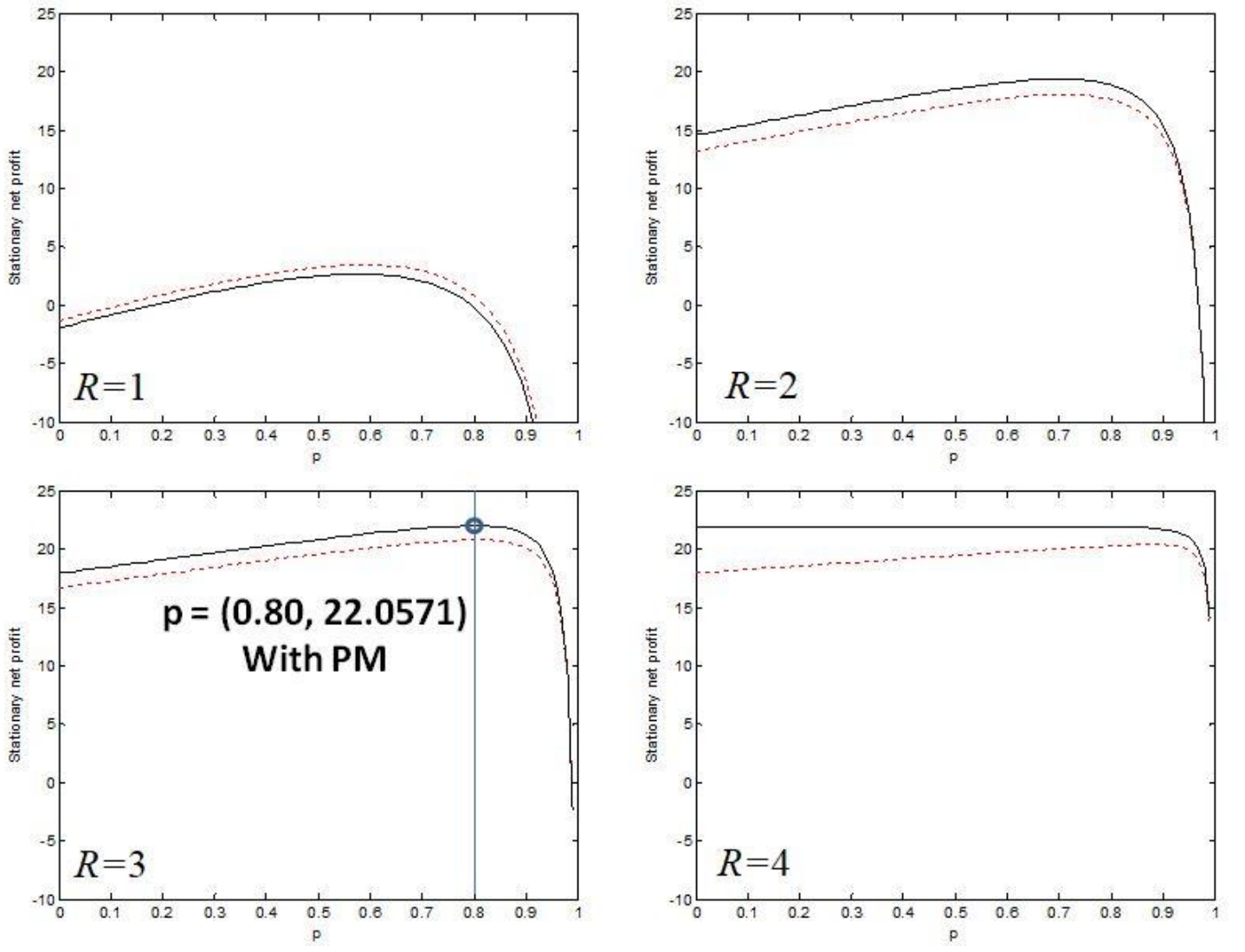

**Figure 4.** Stationary net profit depending on $p$ and $R$ (with preventive maintenance, continuous line; without preventive maintenance, dashed line)

*The Generalized Erlang distribution case*

Analogously to the geometric case, we assume now that the vacation time is distributed as a Generalized Erlang distribution with parameter shape equal to 2. This distribution can be expressed as a phase-type with representation ($\mathbf{v}$, $\mathbf{V}$) being

$$\mathbf{v} = (1, 0) \quad ; \quad \mathbf{V} = \begin{pmatrix} p_1 & 1-p_1 \\ 0 & p_2 \end{pmatrix}.$$

Figures 5 and 6 show the stationary net profit depending on the parameters $p_1$ and $p_2$ and $R$ for the case without preventive maintenance and with preventive maintenance respectively.

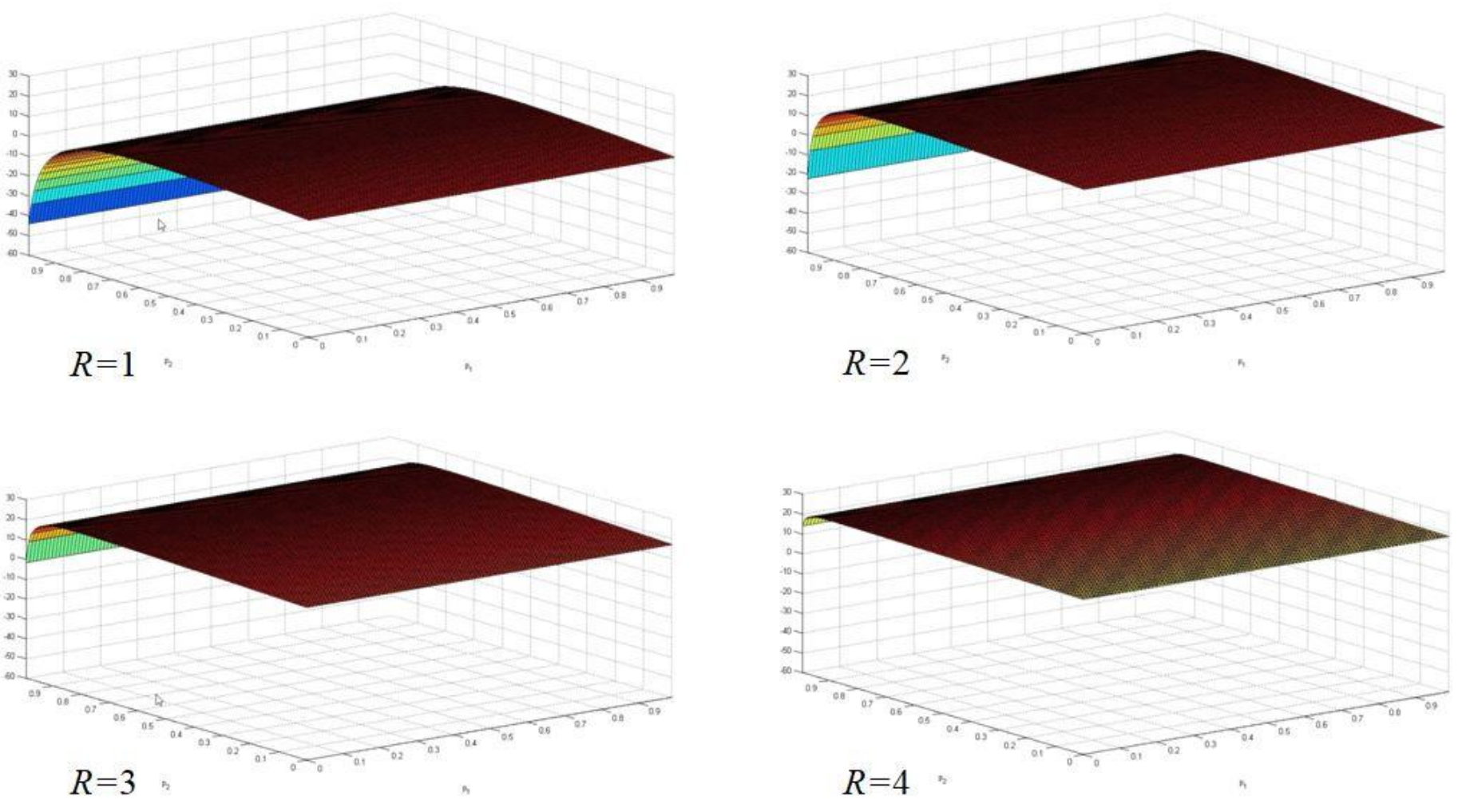

**Figure 5.** Stationary net profit for the system without preventive maintenance depending on *R* and the parameters of the vacation distribution

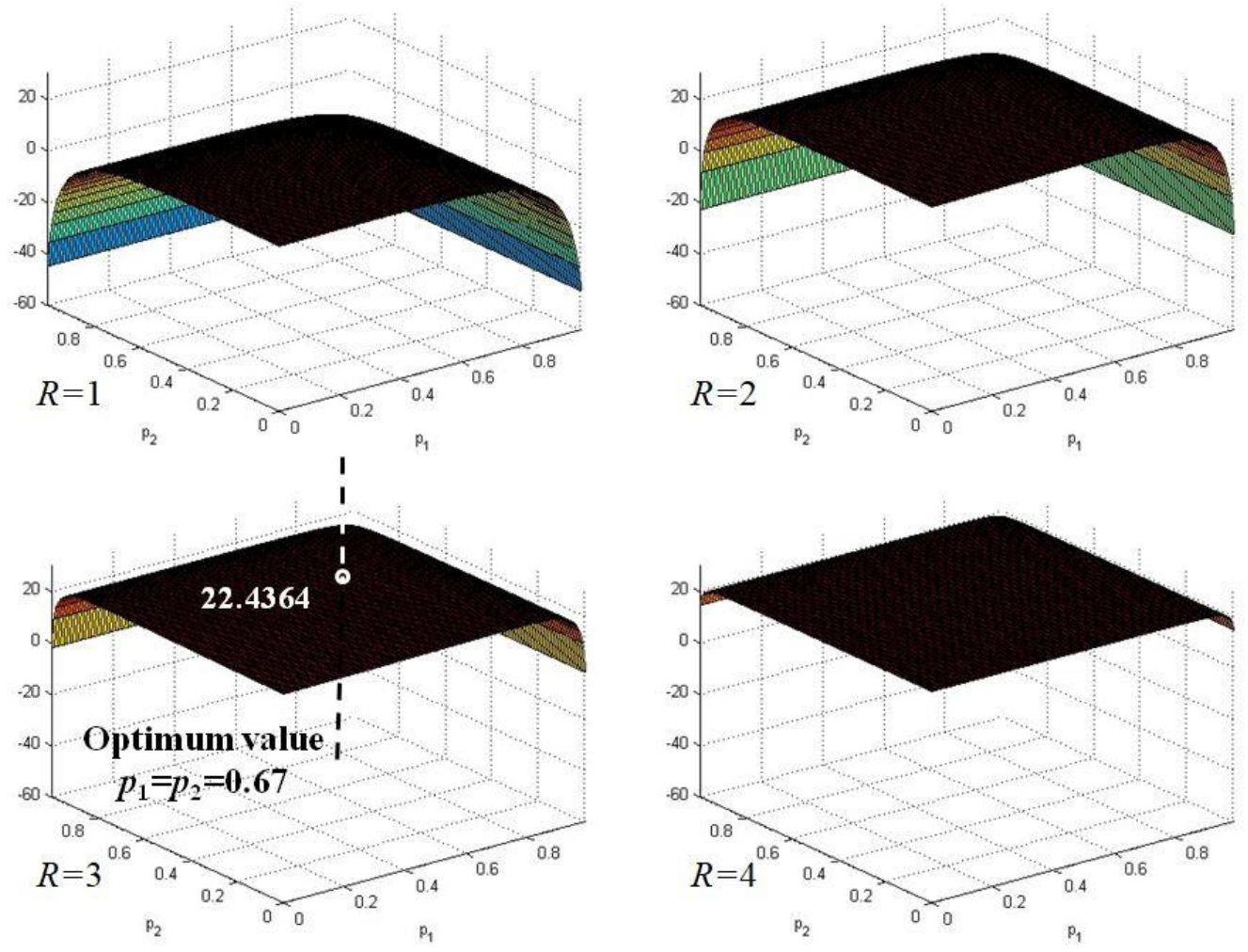

**Figure 6.** Stationary net profit for the system with preventive maintenance depending on *R* and the parameters of the vacation distribution

We can see that, when the generalized Erlang distribution is considered for the vacation time, the optimum value is reached for the preventive maintenance case with $p_1 = p_2 = 0.67$ and $R$=3. In this case, in stationary case, the net profit per unit of time would be equal to 22.4364.

## 6.4. The optimum system with the Generalized Erlang Distribution

In section above we have worked out the optimum system. It is given when the generalized Erlang distribution is considered with parameters (2, 0.67, 0.67) and $R$=3. In this section the performance measures of this system are analysed.

Firstly the time up to first time that the system is replaced (all units undergo a non repairable failure), described in Section 4.3, has been analysed. The reliability function is plotted in Figure 7. Two cases are shown, with and without inspection.

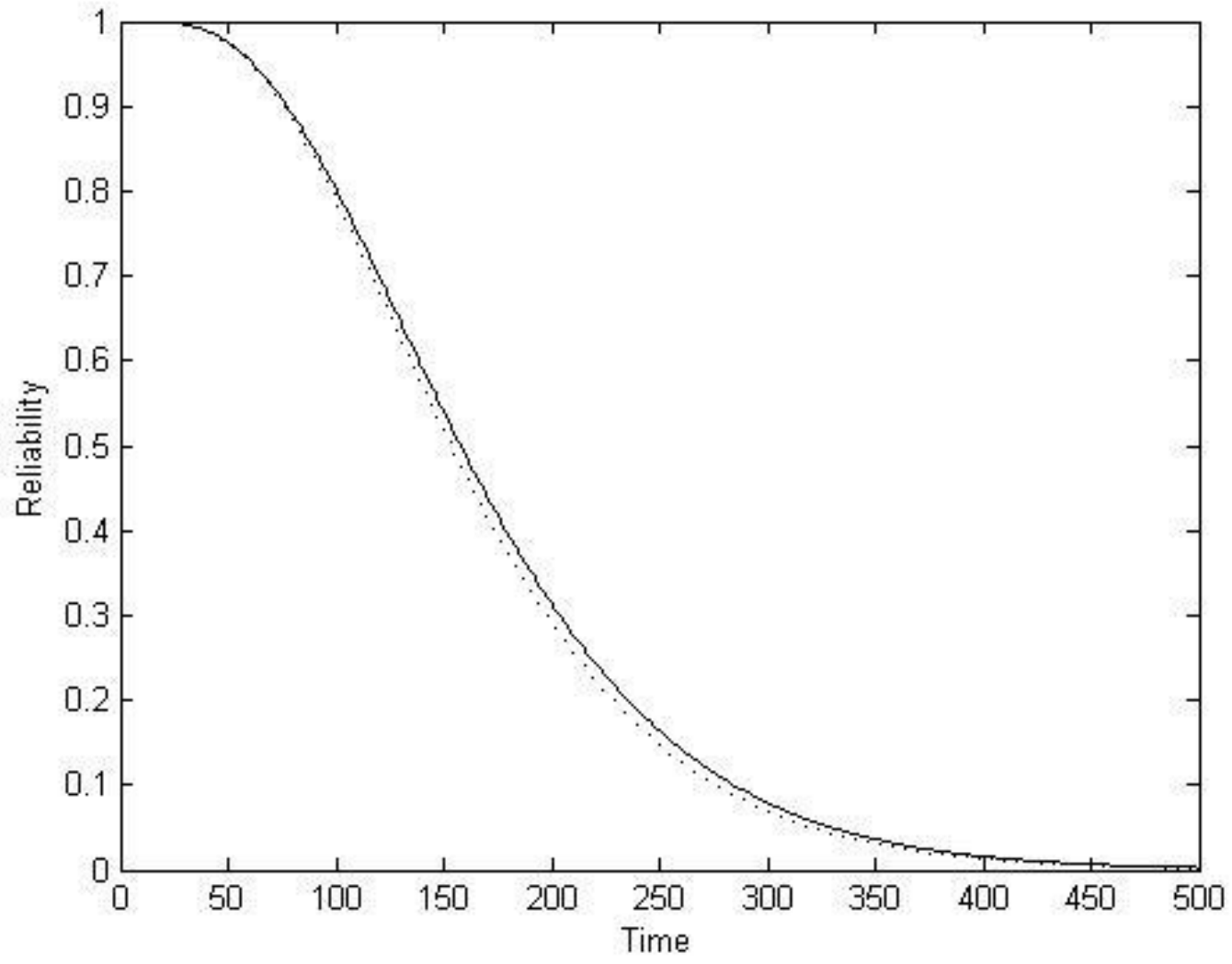

**Figure 7.** Reliability function of the time up to a new system (with inspection, continuous line; without inspection, dashed line)

From the corresponding phase-type distribution, the mean time up to a new system has been calculated in both cases. Thus, the mean time up to be replaced the system for the case without inspection is 167.7631 u.t., and with inspection 172.5269 u.t.

Multiple measures have been achieved for this system with and without inspection. These measures are described in Section 4. Table 2 shows the stationary distribution for macro-states $\mathbf{U}^k$, $k$ units in the system. They can be interpreted as the proportional time that the system is in these macro-states.

**Table 2**. Proportional time in macro-state $\mathbf{U}^k$

| | $\pi_{\mathbf{U}^1}$ | $\pi_{\mathbf{U}^2}$ | $\pi_{\mathbf{U}^3}$ | $\pi_{\mathbf{U}^4}$ |
|---|---|---|---|---|
| Without inspection | 0.3043 | 0.2411 | 0.2306 | 0.2240 |
| With inspection | 0.3057 | 0.2410 | 0.2299 | 0.2234 |

Performance measures are developed for the optimum system with and without inspection following Section 4. Table 3 shows the results.

**Table 3.** Performance measures for the optimum system (without inspection between parentheses)

| $\Upsilon_{nv}$ | $\Upsilon_v$ | $\Upsilon_w$ | $\Upsilon_i$ | $\Lambda^{rep}$ | $\Lambda^{mi}$ | $\Lambda^{NS}$ | $\Phi$ | A |
|---|---|---|---|---|---|---|---|---|
| 0.6806 | 0.3194 | 0.3139 | 0.3667 | 0.0409 | 0.0049 | 0.0058 | 22.4364 | 0.8772 |
| (0.6826) | (0.3174) | (0.3187) | (0.3639) | (0.0432) | | (0.0059) | (21.2077) | (0.8752) |

The proportional time that the repairperson is on vacation is 0.3194. This fact is of interest for the total cost. Therefore, the repairperson is in his workplace a proportion of time of 0.6806 and working a proportion of 0.3139. Then, the 46.12% of the time that the repairperson is in his workplace is working. The remaining time he is idle.

Regarding the mean number of events per unit of time we can observe that it is 0.0409 for repairable failures, 0.0049 for major inspection and 0.0058 for new systems. Thus, each 10000 units of time 58 new systems are expected to be re-started. The availability is also worked out. The 87.72% of the time the system is operational, a 0.23% upper than the without inspection case. Really this is very low but the difference between both net profits is important, 5.79% upper for the case with preventive maintenance.

## 7. Conclusions

Matrix analysis methods can be used to model a complex discrete cold standby system subject to multiple events. This method facilitates the algorithmic and computational development of multi-state complex systems. In the case in question, the online unit within the system is subject to wear and external shocks and may undergo periodic or random inspection. The repair facility is composed of a single repairperson, who may take a vacation (absence) from the repair facility. This repairperson may perform corrective repair and/or preventive maintenance.

The system described is not the standard one in which units are replaced when they undergo a non-repairable failure. In the present study, the analysis takes account of the loss of units following the occurrence of a non-repairable failure. When such a failure occurs, the system continues working with one less unit. This outcome often occurs in practice, and is reflected in the study method presented.

The (indeterminate) number of units within the repair facility and the vacation policy applied determine the behaviour of the repairperson. The vacation time begins when the number of operational units exceeds a given value, and the repairperson will remain in place, without taking a vacation, if the number of operational units in the system is below a pre-determined value.

The system is modelled in an algorithmic and computational form by means of a Markovian Arrival Process with marked arrivals. Matrix-analytic methods are used to obtain the stationary distributions, and multiple measures are derived using a matrix. These measures are related to system performance and financial results.

The method presented in this paper enables us to analyse optimisation problems in multi-state complex systems. A numerical example of such an optimisation is presented. The results obtained show whether preventive maintenance is profitable and reveal the optimum number of operational units, hence determining the appropriate policy for the repairperson's vacation times.

**Author Contributions:** J.E.R.C. was in charge of developing all the statistical theory, computationally implemented the methodology, and obtained the results shown interpretations and conclusions. The only author contributed equally to the writing part of the manuscript.

**Funding:** This paper is partially supported by the project FQM-307 of the Government of Andalusia (Spain) and by the project MTM2017-88708-P of the Spanish Ministry of Science, Innovation and Universities (also supported by the European Regional Development Fund program, ERDF).

**Conflicts of Interest:** The author declares no conflict of interest.

## Appendix A

*Transition probability matrix blocks for the online unit depending on type of event*

$$\mathbf{H}_O = \left(\mathbf{T}\otimes\mathbf{L}+\mathbf{TW}\otimes\mathbf{L}^0\boldsymbol{\gamma}\left(1-\omega^0\right)\right)\otimes\mathbf{M}+\left(\mathbf{U}_1\mathbf{T}\otimes\mathbf{L}+\mathbf{U}_1\mathbf{TW}\otimes\mathbf{L}^0\boldsymbol{\gamma}\left(1-\omega^0\right)\right)\otimes\mathbf{M}^0\boldsymbol{\eta}$$

$$\mathbf{H}_A = \mathbf{T}_r^0\boldsymbol{\alpha}\otimes\mathbf{L}\otimes\mathbf{e}\boldsymbol{\eta}+\left(\mathbf{T}_r^0+\mathbf{TW}_r^0\right)\boldsymbol{\alpha}\otimes\mathbf{L}^0\boldsymbol{\gamma}\left(1-\omega^0\right)\otimes\mathbf{e}\boldsymbol{\eta}$$

$$\mathbf{H'}_A = \mathbf{T}_r^0\otimes\mathbf{L}\otimes\mathbf{e}+\left(\mathbf{T}_r^0+\mathbf{TW}_r^0\right)\otimes\mathbf{L}^0\boldsymbol{\gamma}\left(1-\omega^0\right)\otimes\mathbf{e}$$

$$\mathbf{H}_B = \left[\mathbf{U}_2\left(\mathbf{e}-\mathbf{T}^0\right)\boldsymbol{\alpha}\otimes\mathbf{L}+\mathbf{U}_2\mathbf{T}\left(\mathbf{e}-\mathbf{W}^0\right)\boldsymbol{\alpha}\otimes\mathbf{L}^0\boldsymbol{\gamma}\left(1-\omega^0\right)\right]\otimes\mathbf{M}^0\boldsymbol{\eta}$$

$$\mathbf{H'}_B = \left[\mathbf{U}_2\left(\mathbf{e}-\mathbf{T}^0\right)\otimes\mathbf{L}+\mathbf{U}_2\mathbf{T}\left(\mathbf{e}-\mathbf{W}^0\right)\otimes\mathbf{L}^0\boldsymbol{\gamma}\left(1-\omega^0\right)\right]\otimes\mathbf{M}^0$$

$$\mathbf{H}_C = \left[\mathbf{T}_{nr}^0\boldsymbol{\alpha}\otimes\mathbf{L}+\left(\mathbf{T}_{nr}^0+\mathbf{TW}_{nr}^0\right)\boldsymbol{\alpha}\otimes\mathbf{L}^0\boldsymbol{\gamma}\left(1-\omega^0\right)+\mathbf{e}\boldsymbol{\alpha}\otimes\mathbf{L}^0\boldsymbol{\gamma}\omega^0\right]\otimes\mathbf{e}\boldsymbol{\eta}.$$

$$\mathbf{H'}_C = \left[\mathbf{T}_{nr}^0\otimes\mathbf{L}+\left(\mathbf{T}_{nr}^0+\mathbf{TW}_{nr}^0\right)\otimes\mathbf{L}^0\boldsymbol{\gamma}\left(1-\omega^0\right)+\mathbf{e}\otimes\mathbf{L}^0\boldsymbol{\gamma}\omega^0\right]\otimes\mathbf{e}.$$

## Appendix B

*Matrices for the Markovian Arrival Process depending on the type of event.*

The matrices $\mathbf{D}^A$ and $\mathbf{D}^B$ are developed in the text. The rest are given below.

*The matrix* $\mathbf{D}^O$

The matrix $\boldsymbol{D}^O$ contains the transitions when none event occurs. This matrix is composed of blocks according to the transitions between the macro-states $\mathbf{U}^k$ for $k = 1,\ldots,n$. It is given by

$$
\mathbf{D}^{O}=\begin{pmatrix}
\mathbf{D}_{n}^{O} & & & & \\
 & \mathbf{D}_{n-1}^{O} & & & \\
 & & \mathbf{D}_{n-2}^{O} & & \\
 & & & \ddots & \\
 & & & & \mathbf{D}_{1}^{O}
\end{pmatrix}.
$$

Therefore for the different macro-states it is given by:

- For $k = 1,\ldots, R-1$

$$
\mathbf{D}_{k}^{O}=\begin{array}{c}
E_{0}^{k,nv}\\ E_{1}^{k,nv}\\ \vdots \\ E_{k-1}^{k,vn}\\ E_{k}^{k,nv}
\end{array}
\overset{\begin{array}{ccccc} E_{0}^{k,nv} & E_{1}^{k,nv} & E_{2}^{k,nv} & E_{k-1}^{k,nv} & E_{k}^{k,nv}\end{array}}{\begin{pmatrix}
\mathbf{D}_{00}^{O,k,nv} & & & & \\
\mathbf{D}_{10}^{O,k,nv} & \mathbf{D}_{11}^{O,k,nv} & & & \\
 & \ddots & \ddots & & \\
 & & \mathbf{D}_{k-1,k-2}^{O,k,nv} & \mathbf{D}_{k-1,k-1}^{O,k,nv} & \\
 & & & \mathbf{D}_{k,k-1}^{O,k,nv} & \mathbf{D}_{k,k}^{O,k,nv}
\end{pmatrix}}
$$

- For $k = R,\ldots, n$

$$
\mathbf{D}_{k}^{O}=
\begin{array}{c|cccccccccc|ccccc}
 & E_{0}^{k,v} & E_{1}^{k,v} & \ldots & E_{N-1}^{k,v} & E_{N}^{k,v} & E_{N+1}^{k,v} & \ldots & E_{k-1}^{k,v} & E_{k}^{k,v} & & E_{N}^{k,nv} & E_{N+1}^{k,nv} & E_{N+2}^{k,nv} & E_{k-1}^{k,nv} & E_{k}^{k,nv}\\
\hline
E_{0}^{k,v} & \mathbf{D}_{00}^{O,k,v} & & \ldots & & & & & & & & & & & & \\
E_{1}^{k,v} & & \mathbf{D}_{11}^{O,k,v} & & & & & & & & & & & & & \\
\vdots & & & \ddots & & & & & & & & & & & & \\
E_{N-1}^{k,v} & & & & \mathbf{D}_{N-1,N-1}^{O,k,v} & & & & & & & & & & & \\
E_{N}^{k,v} & & & & & \mathbf{D}_{N,N}^{O,k,v} & & & & & & & & & & \\
E_{N+1}^{k,v} & & & & & & \mathbf{D}_{N+1,N+1}^{O,k,v} & & & & & & & & & \\
\vdots & & & & & & & \ddots & & & & & & & & \\
E_{k-1}^{k,v} & & & & & & & & \mathbf{D}_{k-1,k-1}^{O,k,v} & & & & & & & \\
E_{k}^{k,v} & & & & & & & & & \mathbf{D}_{k,k}^{O,k,v} & & & & & & \\
\hline
E_{N}^{k,nv} & & & & \mathbf{D}_{N,N-1}^{O,k,v} & & & & & & & \mathbf{D}_{N,N}^{O,k,nv} & & & & \\
E_{N+1}^{k,nv} & & & & & & & & & & & \mathbf{D}_{N+1,N}^{O,k,nv} & \mathbf{D}_{N+1,N+1}^{O,k,nv} & & & \\
\vdots & & & & & & & & & & & & \ddots & \ddots & & \\
E_{k-1}^{k,nv} & & & & & & & & & & & & & \mathbf{D}_{k-1,k-2}^{O,k,nv} & \mathbf{D}_{k-1,k-1}^{O,k,nv} & \\
E_{k}^{k,nv} & & & & & & & & & & & & & & \mathbf{D}_{k,k-1}^{O,k,nv} & \mathbf{D}_{k,k}^{O,k,nv}
\end{array}
$$

with

$$\boldsymbol{\theta}=\boldsymbol{\alpha}\otimes\left(\mathbf{L}+\mathbf{L}^{0}\boldsymbol{\gamma}\right)\otimes\boldsymbol{\eta}$$

$$
\mathbf{D}_{N,N-1}^{O,k,v}=\begin{pmatrix}
\mathbf{I}_{2^{N-1}}\otimes\left(I_{\{k=N\}}\boldsymbol{\theta}+I_{\{k\neq N\}}\mathbf{H}_{O}\right)\otimes\mathbf{S}_{1}^{0}\otimes\boldsymbol{\upsilon}\\
\mathbf{I}_{2^{N-1}}\otimes\left(I_{\{k=N\}}\boldsymbol{\theta}+I_{\{k\neq N\}}\mathbf{H}_{O}\right)\otimes\mathbf{S}_{2}^{0}\otimes\boldsymbol{\upsilon}
\end{pmatrix}
$$

$$\mathbf{D}_{r,r}^{O,k,v}=\mathbf{I}_{2^{r}}\otimes\left(I_{\{r<k\}}\mathbf{H}_{O}+I_{\{r=k\}}\left(\mathbf{L}+\mathbf{L}^{0}\boldsymbol{\gamma}\right)\right)\otimes\left(\mathbf{V}+I_{\{r<N\}}\mathbf{V}^{0}\upsilon\right)\quad, r=0,\ldots,k$$

$$\mathbf{D}_{00}^{O,k,nv}=\mathbf{H}_{O}$$

*For* $r = 1,\ldots,k$

$$
\mathbf{D}_{r,r}^{O,k,nv}=\begin{pmatrix}
\mathbf{I}_{2^{r-1}}\otimes\left(I_{\{r<k\}}\mathbf{H}_{O}+I_{\{r=k\}}\left(\mathbf{L}+\mathbf{L}^{0}\boldsymbol{\gamma}\right)\right)\otimes\mathbf{S}_{1} & \mathbf{0}\\
\mathbf{0} & \mathbf{I}_{2^{r-1}}\otimes\left(I_{\{r<k\}}\mathbf{H}_{O}+I_{\{r=k\}}\left(\mathbf{L}+\mathbf{L}^{0}\boldsymbol{\gamma}\right)\right)\otimes\mathbf{S}_{2}
\end{pmatrix}
$$

$$\mathbf{D}_{10}^{O,k,nv}=\begin{pmatrix}\left(I_{\{k>1\}}\mathbf{H}_O+I_{\{k=1\}}\boldsymbol{\theta}\right)\otimes\mathbf{S}_1^0\\ \left(I_{\{k>1\}}\mathbf{H}_O+I_{\{k=1\}}\boldsymbol{\theta}\right)\otimes\mathbf{S}_2^0\end{pmatrix}$$

For $r = 2,\ldots, k$

$$\mathbf{D}_{r,r-1}^{O,k,nv}=\begin{pmatrix}I_{2^{r-2}}\otimes\left(I_{\{r<k\}}\mathbf{H}_O+I_{\{r=k\}}\boldsymbol{\theta}\right)\otimes\mathbf{S}_1^0\otimes\boldsymbol{\beta}_1 & \mathbf{0}\\ \mathbf{0} & I_{2^{r-2}}\otimes\left(I_{\{r<k\}}\mathbf{H}_O+I_{\{r=k\}}\boldsymbol{\theta}\right)\otimes\mathbf{S}_1^0\otimes\boldsymbol{\beta}_2\\ I_{2^{r-2}}\otimes\left(I_{\{r<k\}}\mathbf{H}_O+I_{\{r=k\}}\boldsymbol{\theta}\right)\otimes\mathbf{S}_2^0\otimes\boldsymbol{\beta}_1 & \mathbf{0}\\ \mathbf{0} & I_{2^{r-2}}\otimes\left(I_{\{r<k\}}\mathbf{H}_O+I_{\{r=k\}}\boldsymbol{\theta}\right)\otimes\mathbf{S}_2^0\otimes\boldsymbol{\beta}_2\end{pmatrix}$$

*The matrix* $\mathbf{D}^D$

The matrix $\mathbf{D}^D$ contains the transitions when the repairperson resumes to work without any other event. The structure of this matrix is

$$\mathbf{D}^D=\begin{pmatrix}\mathbf{D}_n^D &&&&&\\ & \mathbf{D}_{n-1}^D &&&&\\ && \ddots &&&\\ &&& \mathbf{D}_R^D &&\\ &&&& \mathbf{0} &\\ &&&&& \ddots &\\ &&&&&& \mathbf{0}\end{pmatrix}.$$

- For $k = R,\ldots, n$

$$\mathbf{D}_k^Y=\begin{array}{c}\\ E_0^{k,v}\\ E_1^{k,v}\\ \vdots\\ E_{N-1}^{k,v}\\ E_N^{k,v}\\ E_{N+1}^{k,v}\\ \vdots\\ E_{k-1}^{k,v}\\ E_k^{k,v}\\ E_N^{k,nv}\\ E_{N+1}^{k,nv}\\ \vdots\\ E_{k-1}^{k,nv}\\ E_k^{k,nv}\end{array}\begin{array}{cccccccccc:ccccc} E_0^{k,v} & E_1^{k,v} & \cdots & E_{N-1}^{k,v} & E_N^{k,v} & E_{N+1}^{k,v} & \cdots & E_{k-1}^{k,v} & E_k^{k,v} & & E_N^{k,nv} & E_{N+1}^{k,nv} & \cdots & E_{k-1}^{k,nv} & E_k^{k,nv}\\ &&&&&&&&&&&&&&\\ &&&&&&&&&&&&&&\\ &&&&&&&&&&&&&&\\ &&&&&&&&&&&&&&\\ &&&&&&&&&& \mathbf{D}_{N,N}^{Y,k,nv} &&&&\\ &&&&&&&&&&& \mathbf{D}_{N+1,N+1}^{Y,k,nv} &&&\\ &&&&&&&&&&&& \ddots &&\\ &&&&&&&&&&&&& \mathbf{D}_{k-1,k-1}^{Y,k,nv} &\\ &&&&&&&&&&&&&& \mathbf{D}_{k,k}^{Y,k,nv}\\ \hdashline &&&&&&&&&&&&&&\\ &&&&&&&&&&&&&&\\ &&&&&&&&&&&&&&\\ &&&&&&&&&&&&&&\\ &&&&&&&&&&&&&&\end{array}$$

For $r =N, \ldots, k$

$$\mathbf{D}_{r,r}^{D,k,nv}=\begin{pmatrix}\mathbf{I}_{2^{r-1}}\otimes\left(I_{\{r=k\}}\left(\mathbf{L}+\mathbf{L}^0\boldsymbol{\gamma}\right)+I_{\{r<k\}}\mathbf{H}_O\right)\otimes\mathbf{V}^0\otimes\boldsymbol{\beta}_1 & \mathbf{0}\\ \mathbf{0} & \mathbf{I}_{2^{r-1}}\otimes\left(I_{\{r=k\}}\left(\mathbf{L}+\mathbf{L}^0\boldsymbol{\gamma}\right)+I_{\{r<k\}}\mathbf{H}_O\right)\otimes\mathbf{V}^0\otimes\boldsymbol{\beta}_2\end{pmatrix}$$

*The matrix* $\mathbf{D}^{AD}$ and $\mathbf{D}^{BD}$

The matrices $\mathbf{D}^{AD}$ and $\mathbf{D}^{BD}$ contain the transitions when the repairperson resumes to work and at same time a repairable failure or major inspection occur. In this case for Y = *AD, BD* we have that

$$\mathbf{D}^Y = \begin{pmatrix} \mathbf{D}_n^Y & & & & & & \\ & \mathbf{D}_{n-1}^Y & & & & & \\ & & \ddots & & & & \\ & & & \mathbf{D}_R^Y & & & \\ & & & & \mathbf{0} & & \\ & & & & & \ddots & \\ & & & & & & \mathbf{0} \end{pmatrix}.$$

- For $k = R, \ldots, n$

$$\mathbf{D}_k^Y = \begin{array}{c|ccccccccc|ccccc} & E_0^{k,v} & E_1^{k,v} & \cdots & E_{N-1}^{k,v} & E_N^{k,v} & E_{N+1}^{k,v} & \cdots & E_{k-1}^{k,v} & E_k^{k,v} & E_N^{k,nv} & E_{N+1}^{k,nv} & \cdots & E_{k-1}^{k,nv} & E_k^{k,nv} \\ \hline E_0^{k,v} & & & & & & & & & & & & & & \\ E_1^{k,v} & & & & & & & & & & & & & & \\ \vdots & & & & & & & & & & & & & & \\ E_{N-1}^{k,v} & & & & & & & & & & \mathbf{D}^{Y,k,nv}_{N-1,N} & & & & \\ E_N^{k,v} & & & & & & & & & & & \mathbf{D}^{Y,k,nv}_{N,N+1} & & & \\ E_{N+1}^{k,v} & & & & & & & & & & & & \ddots & & \\ \vdots & & & & & & & & & & & & & \ddots & \\ E_{k-1}^{k,v} & & & & & & & & & & & & & & \mathbf{D}^{Y,k,nv}_{k-1,k} \\ E_k^{k,v} & & & & & & & & & & & & & & \mathbf{0} \\ \hline E_N^{k,nv} & & & & & & & & & & & & & & \\ E_{N+1}^{k,nv} & & & & & & & & & & & & & & \\ \vdots & & & & & & & & & & & & & & \\ E_{k-1}^{k,nv} & & & & & & & & & & & & & & \\ E_k^{k,nv} & & & & & & & & & & & & & & \end{array}$$

For $r = N-1, \ldots, k-1$

$$\mathbf{F}_{r,r+1}^{AD,k,nv} = \begin{pmatrix} \mathbf{I}_{2^{r-1}} \otimes \left( \left( I_{\{r=k-1\}} \mathbf{H}'_A + I_{\{r<k-1\}} \mathbf{H}_A \right) \otimes \mathbf{V}^0 \otimes \boldsymbol{\beta}_1, \mathbf{0} \right) & \mathbf{0} \\ \mathbf{0} & \mathbf{I}_{2^{r-1}} \otimes \left( \left( I_{\{r=k-1\}} \mathbf{H}'_A + I_{\{r<k-1\}} \mathbf{H}_A \right) \otimes \mathbf{V}^0 \otimes \boldsymbol{\beta}_2, \mathbf{0} \right) \end{pmatrix}$$

$$\mathbf{F}_{r,r+1}^{BD,k,nv} = \begin{pmatrix} \mathbf{I}_{2^{r-1}} \otimes \left( \mathbf{0}, \left( I_{\{r=k-1\}} \mathbf{H}'_B + I_{\{r<k-1\}} \mathbf{H}_B \right) \otimes \mathbf{V}^0 \otimes \boldsymbol{\beta}_1 \right) & \mathbf{0} \\ \mathbf{0} & \mathbf{I}_{2^{r-1}} \otimes \left( \mathbf{0}, \left( I_{\{r=k-1\}} \mathbf{H}'_B + I_{\{r<k-1\}} \mathbf{H}_B \right) \otimes \mathbf{V}^0 \otimes \boldsymbol{\beta}_2 \right) \end{pmatrix}$$

*The matrix* $\mathbf{D}^C$

The matrix $\mathbf{D}^C$ contains the transitions when only a non repairable failure occurs. In this case the matrix is

$$\mathbf{D}^C = \begin{pmatrix} \mathbf{0} & \mathbf{D}_n^C & & & \\ & \mathbf{0} & \mathbf{D}_{n-1}^C & & \\ & & \mathbf{0} & \ddots & \\ & & & \ddots & \mathbf{D}_2^C \\ \mathbf{0} & & & & \mathbf{0} \end{pmatrix}.$$

- For $k = 2, \ldots, R-1$ and $k \neq R \geq 3$

$$
\mathbf{D}_k^C = \begin{array}{c} \\ E_0^{k,nv} \\ E_1^{k,nv} \\ \vdots \\ E_{k-1}^{k,vn} \\ E_k^{k,nv} \end{array}
\begin{array}{c}
\begin{array}{ccccc} E_0^{k-1,nv} & E_1^{k-1,nv} & \ldots & E_{k-2}^{k-1,nv} & E_{k-1}^{k-1,nv} \end{array} \\
\begin{pmatrix}
\mathbf{D}_{00}^{C,k,nv} & & & \\
\mathbf{D}_{10}^{C,k,nv} & \mathbf{D}_{11}^{C,k,nv} & & \\
 & \ddots & \ddots & \\
 & & \mathbf{D}_{k-1,k-2}^{C,k,nv} & \mathbf{D}_{k-1,k-1}^{C,k,nv} \\
 & & & \mathbf{0}
\end{pmatrix}
\end{array}
$$

- For $k = R \geq 2$

$$
\mathbf{D}_k^C = \begin{array}{c} \\ E_0^{k,v} \\ E_1^{k,v} \\ \vdots \\ E_{k-1}^{k,v} \\ E_k^{k,v} \\ E_{N=1}^{k,nv} \\ E_2^{k,nv} \\ \vdots \\ E_{k-1}^{k,nv} \\ E_k^{k,nv} \end{array}
\begin{array}{c}
\begin{array}{ccccc} E_0^{k-1,nv} & E_1^{k-1,nv} & \cdots & E_{k-2}^{k-1,nv} & E_{k-1}^{k-1,nv} \end{array} \\
\left(\begin{array}{ccccc}
\mathbf{0} & & & & \mathbf{0} \\
 & & & & \\
 & \ddots & & & \\
 & & & & \\
\mathbf{0} & & & & \mathbf{0} \\
\hdashline
\mathbf{D}_{10}^{C,k,nv} & \mathbf{D}_{11}^{C,k,nv} & & & \\
 & \mathbf{D}_{21}^{C,k,nv} & \ddots & & \\
 & & \ddots & \ddots & \\
 & & & \mathbf{D}_{k-1,k-2}^{C,k,nv} & \mathbf{D}_{k-1,k-1}^{C,k,nv} \\
\mathbf{0} & \cdots & & \cdots & \mathbf{0}
\end{array}\right)
\end{array}
$$

- For $k = R+1, \ldots, n$ with $R \leq n-1$

$$
\mathbf{D}_k^C = \begin{array}{c} \\ E_0^{k,v} \\ E_1^{k,v} \\ \vdots \\ E_{N-1}^{k,v} \\ E_N^{k,v} \\ E_{N+1}^{k,v} \\ \vdots \\ E_{k-1}^{k,v} \\ E_k^{k,v} \\ E_N^{k,nv} \\ E_{N+1}^{k,nv} \\ \vdots \\ E_{k-1}^{k,nv} \\ E_k^{k,nv} \end{array}
\begin{array}{c}
\begin{array}{cccccccccccccc} E_0^{k-1,v} & E_1^{k-1,v} & \ldots & E_{N-1}^{k-1,v} & E_N^{k-1,v} & E_{N+1}^{k-1,v} & \ldots & E_{k-2}^{k-1,v} & E_{k-1}^{k-1,v} & E_{N-1}^{k-1,nv} & E_N^{k-1,nv} & \ldots & E_{k-2}^{k-1,nv} & E_{k-1}^{k-1,nv} \end{array} \\
\left(\begin{array}{ccccccc:ccccc}
\mathbf{D}_{00}^{C,k,v} & & & & & & & & & & & \\
 & \mathbf{D}_{11}^{C,k,v} & & & & & & & & & & \\
 & & \ddots & & & & & & & & & \\
 & & & \mathbf{D}_{N-1,N-1}^{C,k,v} & & & & & & & & \\
 & & & & \mathbf{D}_{N,N}^{C,k,v} & & & & & & & \\
 & & & & & \mathbf{D}_{N+1,N+1}^{C,k,v} & & & & & & \\
 & & & & & & \ddots & & & & & \\
 & & & & & & \mathbf{D}_{k-1,k-1}^{C,k,v} & & & & & \\
 & & & & & & \mathbf{0} & & & & & \\
\hdashline
 & & & & & & & \mathbf{D}_{N,N-1}^{C,k,nv} & \mathbf{D}_{N,N}^{C,k,nv} & & & \\
 & & & & & & & & \mathbf{D}_{N+1,N}^{C,k,nv} & \mathbf{D}_{N+1,N+1}^{C,k,nv} & & \\
 & & & & & & & & \ddots & \ddots & & \\
 & & & & & & & & & \ddots & \mathbf{D}_{k-1,k-2}^{C,k,nv} & \mathbf{D}_{k-1,k-1}^{C,k,nv} \\
 & & & & & & & & & & & \mathbf{0}
\end{array}\right)
\end{array}
$$

For $r = 0, \ldots, k-1$, $\mathbf{D}_{r,r}^{C,k,v} = \mathbf{I}_{2^r} \otimes \left( I_{\{r=k-1\}} \mathbf{H}'_C + I_{\{r<k-1\}} \mathbf{H}_C \right) \otimes \left( \mathbf{V} + I_{\{r<N-1\}} \mathbf{V}^0 \upsilon \right)$

$\mathbf{D}_{00}^{C,k,nv} = \mathbf{H}_C$ ;

For $r = 1, \ldots, k-1$ ;

$$
\mathbf{D}_{r,r}^{C,k,nv} = \begin{pmatrix} \mathbf{I}_{2^{r-1}} \otimes \left( I_{\{r=k-1\}} \mathbf{H}'_C + I_{\{r<k-1\}} \mathbf{H}_C \right) \otimes \mathbf{S}_1 & \mathbf{0} \\ \mathbf{0} & \mathbf{I}_{2^{r-1}} \otimes \left( I_{\{r=k-1\}} \mathbf{H}'_C + I_{\{r<k-1\}} \mathbf{H}_C \right) \otimes \mathbf{S}_2 \end{pmatrix}
$$

$$
\mathbf{D}_{10}^{C,k,nv} = \begin{pmatrix} \mathbf{H}_C \otimes \mathbf{S}_1^0 \\ \mathbf{H}_C \otimes \mathbf{S}_2^0 \end{pmatrix}
$$

For $r = 2,\ldots,k-1$, $\mathbf{D}_{r,r-1}^{C,k,nv} = \begin{pmatrix} I_{2^{r-2}} \otimes \mathbf{H}_C \otimes \mathbf{S}_1^0 \otimes \boldsymbol{\beta}_1 & \mathbf{0} \\ \mathbf{0} & I_{2^{r-2}} \otimes \mathbf{H}_C \otimes \mathbf{S}_1^0 \otimes \boldsymbol{\beta}_2 \\ I_{2^{r-2}} \otimes \mathbf{H}_C \otimes \mathbf{S}_2^0 \otimes \boldsymbol{\beta}_1 & \mathbf{0} \\ \mathbf{0} & I_{2^{r-2}} \otimes \mathbf{H}_C \otimes \mathbf{S}_2^0 \otimes \boldsymbol{\beta}_2 \end{pmatrix}$

*The matrix* $\mathbf{D}^{CD}$

The matrix $\mathbf{D}^{CD}$ contains the transitions when a non repairable failure occurs and the repairperson resumes to his work. In this case the matrix is

$$\mathbf{D}^{CD} = \begin{pmatrix} \mathbf{0} & \mathbf{D}_n^{CD} & & & & & & \\ & \ddots & \ddots & & & & & \\ & & \mathbf{0} & \mathbf{D}_R^{CD} & & & & \\ & & & \mathbf{0} & \mathbf{0} & & & \\ & & & & \ddots & \ddots & & \\ & & & & & \mathbf{0} & \mathbf{0} & \\ & & & & & & \mathbf{0} & \end{pmatrix}$$

- For $k = R$

$$\mathbf{D}_k^{CD} = \begin{array}{c} \\ E_0^{k,v} \\ E_1^{k,v} \\ \vdots \\ E_{k-1}^{k,v} \\ E_k^{k,v} \\ E_{N=1}^{k,nv} \\ E_2^{k,nv} \\ \vdots \\ E_{k-1}^{k,nv} \\ E_k^{k,nv} \end{array} \begin{array}{c} \begin{array}{ccccc} E_0^{k-1,nv} & E_1^{k-1,nv} & \cdots & E_{k-2}^{k-1,nv} & E_{k-1}^{k-1,nv} \end{array} \\ \left(\begin{array}{cccc} \mathbf{D}_{00}^{CD,k,nv} & & & \mathbf{0} \\ & \mathbf{D}_{11}^{CD,k,nv} & & \\ & & \ddots & \\ & & & \mathbf{D}_{k-1,k-1}^{CD,k,nv} \\ \mathbf{0} & & & \mathbf{0} \\ \hdashline \mathbf{0} & \cdots & \cdots & \mathbf{0} \\ & & & \\ & & & \\ & & & \\ \mathbf{0} & \cdots & \cdots & \mathbf{0} \end{array}\right) \end{array}$$

The matrix blocks for the case $k = R$ are

$$\mathbf{D}_{00}^{CD,k,nv} = \mathbf{H}_C \otimes \mathbf{e}$$

For $r = 1, \ldots, k-1$

$$\mathbf{D}_{r,r}^{CD,k,nv} = \begin{pmatrix} \mathbf{I}_{2^{r-1}} \otimes \left( I_{\{r=k-1\}} \mathbf{H}'_C + I_{\{r<k-1\}} \mathbf{H}_C \right) \otimes \mathbf{e} \otimes \boldsymbol{\beta}_1 & \mathbf{0} \\ \mathbf{0} & \mathbf{I}_{2^{r-1}} \otimes \left( I_{\{r=k-1\}} \mathbf{H}'_C + I_{\{r<k-1\}} \mathbf{H}_C \right) \otimes \mathbf{e} \otimes \boldsymbol{\beta}_2 \end{pmatrix}$$

- For $k = R+1,\ldots, n$ and $R \le n-1$

$$
\mathbf{D}_k^{CD} = \begin{array}{c} \\ E_0^{k,v} \\ E_1^{k,v} \\ \vdots \\ E_{N-1}^{k,v} \\ E_N^{k,v} \\ E_{N+1}^{k,v} \\ \vdots \\ E_{k-1}^{k,v} \\ E_k^{k,v} \\ E_N^{k,nv} \\ E_{N+1}^{k,nv} \\ \vdots \\ E_{k-1}^{k,nv} \\ E_k^{k,nv} \end{array}
\begin{array}{c} E_0^{k-1,v}\ E_1^{k-1,v} \cdots E_{N-1}^{k-1,v}\ E_N^{k-1,v}\ E_{N+1}^{k-1,v} \cdots E_{k-2}^{k-1,v}\ E_{k-1}^{k-1,nv} \quad E_{N-1}^{k-1,nv}\ E_N^{k-1,nv} \cdots E_{k-2}^{k-1,nv}\ E_{k-1}^{k-1,nv} \\
\left(\begin{array}{c:cccc}
 & & & & \\
 & \mathbf{D}_{N-1,N-1}^{CD,k,nv} & & & \\
 & & \mathbf{D}_{N,N}^{CD,k,nv} & & \\
 & & & \ddots & \\
 & & & & \mathbf{D}_{k-1,k-1}^{CD,k,nv} \\
 & & & & \mathbf{0} \\ \hdashline
 & & & & \\
 & & & &
\end{array}\right)
\end{array}
$$

- The matrix blocks for the case $k = R+1, \ldots, n$ are

For $r = N-1, \ldots, k-1$

$$
\mathbf{D}_{r,r}^{CD,k,nv} = \begin{pmatrix} \mathbf{I}_{2^{r-1}} \otimes \left( I_{\{r<k-1\}} \mathbf{H}_C + I_{\{r=k-1\}} \mathbf{H}_C^{'} \right) \otimes \mathbf{V}^0 \otimes \boldsymbol{\beta}_1 & \mathbf{0} \\ \mathbf{0} & \mathbf{I}_{2^{r-1}} \otimes \left( I_{\{r<k-1\}} \mathbf{H}_C + I_{\{r=k-1\}} \mathbf{H}_C^{'} \right) \otimes \mathbf{V}^0 \otimes \boldsymbol{\beta}_2 \end{pmatrix}
$$

*The matrix* $\mathbf{D}^{NS}$

The matrix $\mathbf{D}^{NS}$ contains the transitions when a failure provokes the system is restarted. Obviously, in this case the system is composed of only one unit. When this one is broken, a new system with $n$ units re-starts. When it occurs, the vacation time begins again. The structure of the matrix is

$$
\mathbf{D}^{NS} = \begin{pmatrix} \mathbf{0} & & & & \\ & \mathbf{0} & & & \\ & & \mathbf{0} & & \\ & & & \ddots & \\ \mathbf{D}_1^{NS} & & & & \mathbf{0} \end{pmatrix}.
$$

- If $R = 1$

$$
\mathbf{D}_1^{NS} = \begin{array}{c} \\ E_0^{1,v} \\ E_1^{1,v} \\ E_1^{1,nv} \end{array}
\begin{array}{c} E_0^{n,v}\ E_1^{n,v} \cdots E_{N-1}^{n,v}\ E_N^{n,v}\ E_{N+1}^{n,v} \cdots E_{k-2}^{n,v}\ E_{k-1}^{n,nv}\ E_N^{n,nv}\ E_{N+1}^{n,nv} \cdots E_{k-2}^{n,nv}\ E_{k-1}^{n,nv} \\
\begin{pmatrix} \mathbf{D}_{00}^{NS,1,v} & \mathbf{0} & \cdots & \cdots & \mathbf{0} \\ \mathbf{0} & \mathbf{0} & \cdots & \cdots & \mathbf{0} \\ \mathbf{0} & \mathbf{0} & \cdots & & \mathbf{0} \end{pmatrix}
\end{array}
$$

with $\mathbf{D}_{00}^{NS,1,v} = \mathbf{H}_C \otimes \mathbf{e}_\upsilon \mathbf{v}$.

- If $R > 1$

$$
\mathbf{D}_1^{NS} = \begin{array}{c} \\ E_0^{1,nv} \\ E_1^{1,nv} \end{array}
\begin{array}{c} E_0^{n,v}\ E_1^{n,v} \cdots E_{N-1}^{n,v}\ E_N^{n,v}\ E_{N+1}^{n,v} \cdots E_k^{n,v}\ E_1^{n,nv}\ E_2^{n,nv}\ E_3^{n,nv} \cdots E_{n-1}^{n,nv}\ E_n^{n,nv} \\
\begin{pmatrix} \mathbf{D}_{00}^{NS,1,v} & \mathbf{0} & \cdots & \cdots & \mathbf{0} \\ \mathbf{0} & \mathbf{0} & \cdots & \cdots & \mathbf{0} \end{pmatrix}
\end{array}
$$

with $\mathbf{D}_{00}^{NS,1,v} = \mathbf{H}_C \otimes \mathbf{v}$.

**Appendix C**

To calculate the expected times that the repairperson returns to workplace, independently of he remains or begins another period of vacation, the following matrix **Q** is defined. This matrix is built analogously to the matrix **D**, but any return is considered. Therefore, the matrix **Q** is the addition of the following matrices

$$\mathbf{Q} = \mathbf{D}^{O}_{r-b} + \mathbf{D}^{A}_{r-b} + \mathbf{D}^{B}_{r-b} + \mathbf{D}^{C}_{r-b} + \mathbf{D}^{D} + \mathbf{D}^{AD} + \mathbf{D}^{BD} + \mathbf{D}^{CD} + \mathbf{D}^{NS}_{r-b}.$$

The matrices $\mathbf{D}^{D}, \mathbf{D}^{AD}, \mathbf{D}^{BD}, \mathbf{D}^{CD}$ are described in Appendix B. The other matrices have the same structure for the corresponding event given in Appendix B. These matrices are of zeros excepting the following blocks.

- For $r = 0, \ldots, k-R$ and $k \geq R$

  $$\mathbf{D}^{O,k,v}_{r,r} = \mathbf{I}_{2^r} \otimes \mathbf{H}_O \otimes \mathbf{V}^0 \upsilon$$
- For $r = 1, \ldots, k-R-1$ and $k \geq R+2$

  $$\mathbf{D}^{A,k,v}_{0,1} = \left(\mathbf{H}_A \otimes \mathbf{V}^0 \mathbf{v}, \mathbf{0}\right); \quad \mathbf{D}^{B,k,v}_{0,1} = \left(\mathbf{0}, \mathbf{H}_B \otimes \mathbf{V}^0 \mathbf{v}\right)$$

  $$\mathbf{D}^{A,k,v}_{r,r+1} = \mathbf{I}_{2^r} \otimes \left(\mathbf{H}_A \otimes \mathbf{V}^0 \mathbf{v}, \mathbf{0}\right)$$

  $$\mathbf{D}^{B,k,v}_{r,r+1} = \mathbf{I}_{2^r} \otimes \left(\mathbf{0}, \mathbf{H}_B \otimes \mathbf{V}^0 \mathbf{v}\right).$$
- For $r = 0, \ldots, k-R-1$ and $k \geq R+1$

  $$\mathbf{D}^{C,k,v}_{r,r} = \mathbf{I}_{2^r} \otimes \mathbf{H}_C \otimes \mathbf{V}^0 \upsilon$$
- If $R=1$, $\mathbf{D}^{NS,1,v}_{00} = \mathbf{H}_C \otimes \mathbf{V}^0 \mathbf{v}$